\documentclass[10pt,english,allowdisplaybreaks,onecolumn,draftcls]{IEEEtran}
\usepackage[T1]{fontenc}
\usepackage[latin1]{inputenc}
\usepackage{geometry}
\usepackage{color}
\usepackage{float}
\usepackage{amsthm}
\usepackage{amsmath}
\usepackage{amssymb}
\usepackage{graphicx}
\usepackage{setspace}
\usepackage{esint}
\makeatletter

\providecommand{\tabularnewline}{\\}

\theoremstyle{plain}
\newtheorem{thm}{\protect\theoremname}
\theoremstyle{definition}
\newtheorem{defn}{\protect\definitionname}
\theoremstyle{plain}
\newtheorem{lem}{\protect\lemmaname}
\theoremstyle{plain}
\newtheorem{cor}{\protect\corollaryname}

\@ifundefined{definecolor}
 {\usepackage{color}}{}
\@ifundefined{definecolor}{\usepackage{color}}{}

\providecommand{\theoremname}{Theorem}

\allowdisplaybreaks
\usepackage{babel}
\providecommand{\corollaryname}{Corollary}
\providecommand{\definitionname}{Definition}
\providecommand{\lemmaname}{Lemma}
\providecommand{\theoremname}{Theorem}

\providecommand{\corollaryname}{\inputencoding{latin9}Corollary}
\providecommand{\definitionname}{\inputencoding{latin9}Definition}
\providecommand{\lemmaname}{\inputencoding{latin9}Lemma}
\providecommand{\theoremname}{\inputencoding{latin9}Theorem}

\makeatother

\providecommand{\corollaryname}{\inputencoding{latin9}Corollary}
\providecommand{\definitionname}{\inputencoding{latin9}Definition}
\providecommand{\lemmaname}{\inputencoding{latin9}Lemma}
\providecommand{\theoremname}{\inputencoding{latin9}Theorem}

\begin{document}

\title{Nested Lattice Codes for Gaussian Two-Way Relay Channels}

\author{\authorblockA{Shahab Ghasemi-Goojani and Hamid Behroozi\\
 Information Systems and Security Lab (ISSL)\\
 Department of Electrical Engineering\\
 Sharif University of Technology\\
 Tehran, Iran\\
 Email:shahab\_ghasemi@ee.sharif.edu, behroozi@sharif.edu}}
\maketitle
\begin{abstract}
In this paper, we consider a Gaussian two-way relay channel (GTRC),
where two sources exchange messages with each other through a relay.
We assume that there is no direct link between sources, and all nodes
operate in full-duplex mode. By utilizing nested lattice codes for
the uplink (i.e., MAC phase), and structured binning for the downlink
(i.e., broadcast phase), we propose two achievable schemes. Scheme
1 is based on ``compute and forward'' scheme of \cite{Nazar_IT_11}
while scheme 2 utilizes two different lattices for source nodes based
on a three-stage lattice partition chain. We show that scheme 2 can
achieve capacity region at the high signal-to-noise ratio (SNR). Regardless
all channel parameters, the achievable rate of scheme 2 is within
0.2654 bit from the cut-set outer bound for user 1. For user 2, the
proposed scheme achieves within 0.167 bit from the outer bound if
channel coefficient is larger than one, and achieves within 0.2658
bit from the outer bound if channel coefficient is smaller than one.
Moreover, sum rate of the proposed scheme is within $0.334$ bits
from the sum capacity. These gaps for GTRC are the best gap-to-capacity
results to date. 
\end{abstract}

\section{Introduction}

In this paper, we study a two-way relay channel, where two nodes exchange
their messages with each other via a relay. This can be considered
as, e.g., two mobile users communicate to each other via the access
point in a WLAN \cite{Zhao_thesis_ETH_2010} or a satellite which
enables data exchange between several earth stations where there is
no direct link between the stations.

Two-way or bi-directional communication between two nodes without
a relay was first proposed and studied by Shannon in \cite{Shannon_61}.
In this setup, two nodes want to exchange messages with each other,
and act as transmitters and receivers at the same time. For this setup,
the capacity region is not known in general and only inner and outer
bounds on the capacity region are obtained in the literature.

The two-way relay channel, also known as the bi-directional relay
channel, consists of two nodes communicating with each other in a
bi-directional manner via a relay. This setup was first introduced
in \cite{Rankovi_Allerton_05}; later studied in \cite{Rankovi_Allerton_05,Gunduz_Allerton_2008,Avestimehr_ETT_2010}
and an approximate characterization of the capacity region of the
Gaussian case is derived.

The traditional relaying protocols require four channel uses to exchange
the data of two nodes whereas the two-way relaying protocol in \cite{Rankovi_Allerton_05}
only needs two phases to achieve bidirectional communication between
the two nodes. The first phase is referred to as the multiple access
(MAC) phase, and the second phase is referred to as the broadcast
(BRC) phase. In the MAC phase, both nodes transmit their messages
to the relay node which decodes them. In the BRC phase, the relay
combines the data from both nodes and broadcasts the combined data
back to both nodes. For this phase, there exist several strategies
for the processing at the relay node, e.g., an amplify-and-forward
(AF) strategy \cite{Rankovi_Allerton_05}, a decode-and-forward (DF)
strategy \cite{Rankovi_Allerton_05,Oechtering_ISITA_06}, or a compress-and-forward
(CF) strategy \cite{Schnurr_Asilomar_07}.

The AF protocol is a simple scheme, which amplifies the signal transmitted
from both nodes and retransmit it to them, and unlike the DF protocol,
no decoding at the relay is performed. In the two-way AF relaying
strategy, the signals at the relay are actually combined on the symbol
level. Due to amplification of noise, its performance degrades at
low signal-to-noise ratios (SNR). The two-way DF relaying strategy
was proposed in \cite{Rankovi_Allerton_05}, where the relay decodes
the received data bits from the both nodes. Since the decoded data
at the relay can be combined on the symbol level or on the bit level,
there has been different data combining schemes at the relay for the
two-way DF relaying strategy: superposition coding, network coding,
and lattice coding \cite{Zhao_thesis_ETH_2010}. In the superposition
coding scheme, applied in \cite{Rankovi_Allerton_05}, the data from
the two nodes are combined on the symbol level, where the relay sends
the linear sum of the decoded symbols from both nodes. Shortly after
the introduction of the two-way relay channel, its connection to network
coding \cite{Ahlswede_IT_00} was observed and investigated. The network
coding schemes combine the data from nodes on the bit level using
the XOR operation, see e.g., \cite{Wu_CISS_05,Larsson_06_VTC,Popovski_07,Zhang_JSAC_2009,Liu_09_Globecom,Zhou_10}.
Lattice coding uses modulo addition in a multi-dimensional space and
utilizes nonlinear operations for combining the data. Applying lattice
coding in two-way relaying systems was considered in, e.g., \cite{Baik_08,Nam_IT_2010,Wilson_IT_2010_Relay}.
In general, as in CF or partial DF relaying strategies, the relay
node need not to decode the source messages, but only need to pass
sufficient information to the destination nodes.

In this paper, we focus on the Gaussian TRC (GTRC) as shown in Fig.
\ref{fig:System-Model}, where nodes 1 and 2 want to communicate with
each other at rates $R_{1}$ and $R_{2}$, respectively, while a relay
node facilitates the communication between them. TRC with a direct
link between nodes 1 and 2 is studied in \cite{Kim_IT_08,Kim_IT_11,Ashar_Allerton_2012,Zhong_ICC_2012,Tian_TWC_12}.
Here, we consider a GTRC without a direct link between nodes 1 and
2. Our model is an extension of \cite{Wilson_IT_2010_Relay} and it
is essentially the same as those considered in \cite{Gunduz_Allerton_2008,Avestimehr_Allerton_2008,Nam_IT_2010,Nam_IZSC_2008,Knopp_GDR_2007,Ong_ISIT_11}.
Similar to \cite{Wilson_IT_2010_Relay,Nam_IT_2010}, we apply the
nested lattice coding. For a comprehensive review on lattices and
their performance analysis, we refer the reader to the study of \cite{Erez_IT_05,Zamir_09_ITA,Forney_Allerton_2003}.
Nested lattice codes have been shown to be capacity achieving for
the AWGN channel \cite{Erez_IT_04,Zamir_02}, AWGN broadcast channel
\cite{Zamir_02} and the AWGN multiple access channel \cite{Nazar_IT_11}.
Song and Devroye \cite{Song_sub}, using list decoding technique,
show that lattice codes can achieve the DF rate of single source,
single destination Gaussian relay channels with one or more relays.
It is also shown that the lattice CF scheme, which exploits a lattice
Wyner-Ziv binning, for the Gaussian relay channel achieves the same
rate as the Cover-El Gamal CF achievable rate \cite{Song_sub}. Nazer
and Gastpar by introducing compute-and-forward scheme obtain significantly
higher rates between users than previous results in the literature
over a relay network \cite{Nazar_IT_11}. By utilizing a deterministic
approach \cite{Avestimehr_Allerton_2007} and a simple AF or a particular
superposition coding strategy at the relay, \cite{Avestimehr_ETT_2010}
achieves full-duplex GTRC capacity region within 3 bits from the cut-set
bound for all values of channel gains. A general GTRC is considered
in \cite{Nam_IT_2010}, and it is shown that regardless of all channel
parameters, the capacity region for each user is achievable within
$\frac{1}{2}$ bit and the sum rate is within $\log\frac{3}{2}$ bits
from the sum capacity, which was the best gap-to-capacity. Communication
over a symmetric GTRC when all source and relay nodes have the same
transmit powers and noise variances, is considered in \cite{Wilson_IT_2010}.
It is shown that lattice codes and lattice decoding at the relay node
can achieve capacity region at the high SNR.\textcolor{red}{{} }\textcolor{black}{Lim
et.al, using noisy network coding for GTRC, show that the achievable
rate of their scheme for each user is within $\frac{1}{2}$ bit from
the cut-set bound.} \cite{Lim_IT_11}. The achievable sum rate in
\cite{Lim_IT_11} is within 1 bit from the cut-set bound.

Here, we apply nested lattice codes at the relay to obtain linear
combination of messages. By comparing with the cut-set outer bound,
we show that the achievable rate region, regardless of all channel
parameters, is within 0.2654 bit from the outer bound for user 1.
For user 2, the proposed scheme achieves within 0.167 bit from the
outer bound if the channel coefficient is larger than one, and achieves
within 0.2658 bit from the outer bound if channel coefficient $\left(g\right)$
is smaller than one. The achievable sum rate by the proposed scheme
is within $0.334$ bit from the cut-set bound. Thus, the proposed
scheme outperforms \cite{Nam_IT_2010,Avestimehr_ETT_2010,Lim_IT_11}
and the resulting gap is the best gap-to-capacity result to date.

The remainder of the paper is organized as follows. We present the
preliminaries of lattice codes and the channel model in Section \ref{sec:Preliminaries:-Lattices-and}.
In Section \ref{sec:Lattice-based-achievable-rate}, we introduce
two achievable coding schemes. Scheme 1 is based on compute-and-forward
strategy while scheme 2 utilizes two different lattices for source
nodes based on a three-stage lattice partition chain. In Section \ref{sec:Outer-Bound},
we analyze the gap between the achievable rate region and the cut-set
outer bound for each user. Numerical results and performance comparison
between two schemes and the outer bound are presented in Section \ref{sec:Numerical-Result}.
Section \ref{sec:Conclusion} concludes the paper.

\section{\label{sec:Preliminaries:-Lattices-and}Preliminaries: Lattices and
Channel Model}

\subsection{Notations and Channel Model}

Throughout the paper, random variables and their realizations are
denoted by capital and small letters, respectively. $\boldsymbol{x}$
stands for a vector of length $n$, $(x_{1},x_{2},...,x_{n})$. Also,
$\left\Vert \boldsymbol{.}\right\Vert $ denotes the Euclidean norm,
and all logarithms are with respect to base $2$.

\begin{figure}
\begin{centering}
\includegraphics[width=12cm]{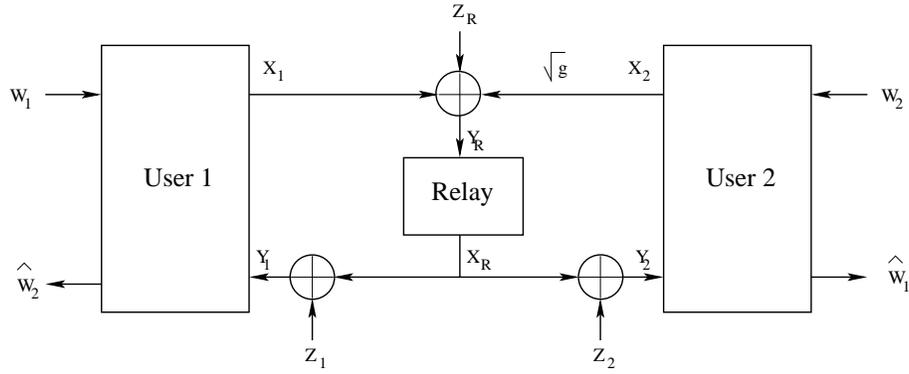} 
\par\end{centering}

\caption{\label{fig:System-Model}System Model. Two communication phases for
a Gaussian two-way relay channel: MAC phase and Broadcast phase.}
\end{figure}

In this paper, we consider a Gaussian two-way relay channel (GTRC),
with two sources that exchange messages through a relay. We assume
that there is no direct link between the sources and all nodes operate
in a full-duplex mode. The system model is depicted in Fig. \ref{fig:System-Model}.
Communication process takes place in two phases: MAC phase BRC phase,
which are described in the following: 
\begin{itemize}
\item \textbf{MAC phase}: In this phase, first the input message to both
encoders, $W_{1},W_{2}$, are mapped to 
\[
X_{i}^{(t)}=f_{i}(W_{i},Y_{i}^{t-1}),\,\,\,\,\,\,\textrm{f}\textrm{or}\,\,\, i=1,2
\]
 where $f_{i}$ is the encoder function at node $i$ and $Y_{i}^{t-1}=\left\{ Y_{i}^{(1)},Y_{i}^{(2)},\cdots,Y_{i}^{(t-1)}\right\} $
is the set of past channel outputs at node $i$. Without loss of generality,
we assume that both transmitted sequences $X_{1}^{(t)}$, $X_{2}^{(t)}$
are average-power limited to $P>0$, i.e., 
\begin{equation}
\frac{1}{n}\sum_{t=1}^{n}\mathbb{E}\left[\left|X_{i}^{(t)}\right|^{2}\right]\leq P,\,\,\,\,\,\textrm{f}\textrm{or}\,\,\, i=1,2.\label{PowerConstraint}
\end{equation}
 Both nodes send their signals to the relay. The received signal at
the relay at time $t$ is specified by 
\begin{eqnarray*}
Y_{R}^{(t)} & = & X_{1}^{(t)}+\sqrt{g}X_{2}^{(t)}+Z_{R}^{(t)},
\end{eqnarray*}
 where $X_{1}^{(t)}$ and $X_{2}^{(t)}$ are the signals transmitted
from node 1 and node 2 at time $t$, respectively. $\sqrt{g}$ denotes
the channel gain between node 2 and the relay and all other channel
gains are assumed to be one. $Z_{R}^{(t)}$ represents an independent
identically distributed (i.i.d.) Gaussian random variable with mean
zero and variance $N_{R}$, which models an additive white Gaussian
noise (AWGN) at the relay. 
\item \textbf{BRC phase}: During the broadcast phase, the relay node processes
the received signal and retransmits the combined signals back to both
nodes, i.e., the relay communicates signal $X_{R}$ to both nodes
1 and 2. Since the relay has no messages of its own, the relay signal
at time $t$, $X_{R}^{(t)}$, is a function of the past relay inputs,
i.e., 
\[
X_{R}^{(t)}=f_{R}\left(Y_{R}^{t-1}\right),
\]
 where $f_{R}$ is a encoding function at the relay and $Y_{R}^{t-1}=\left\{ Y_{R}^{(1)},Y_{R}^{(2)},\cdots,Y_{R}^{(t-1)}\right\} $is
a sequence of past relay inputs. The power constraint at the relay
is given by 
\[
\frac{1}{n}\sum_{t=1}^{n}\mathbb{E}\left[\left|X_{R}^{(t)}\right|^{2}\right]\leq P_{R}.
\]
 The received signal at each node at time $t$ is given by 
\[
Y_{i}^{(t)}=X_{R}^{(t)}+Z_{i}^{(t)},\,\,\,\,\,\,\textrm{f}\textrm{or}\,\,\, i=1,2
\]
 where $X_{R}^{(t)}$ is the transmitted signal from the relay node
at time $t$ and $Z_{i}^{(t)}$ represents an i.i.d AWGN with zero
mean and variance $N_{i}$. 
\end{itemize}
Node 1, based on the received sequence, $\boldsymbol{Y}_{1}$, and
its own message, $W_{1}$, makes an estimate of the other message,
$W_{2}$, as 
\[
\hat{W_{2}}=\psi_{1}\left(\boldsymbol{Y}_{1},W_{1}\right),
\]
 where $\psi_{1}$ is a decoding function for at node 1. Decoding
at node 2 is performed in a similar way. The average probability of
error is defined as 
\[
P_{e}=\textrm{ Pr}\left\{ \hat{W}_{1}\neq W_{1}\,\,\textrm{or }\,\,\hat{W}_{2}\neq W_{2}\right\} .
\]
 A rate pair $\left(R_{1},R_{2}\right)$ of non-negative real values
is achievable if there exists encoding and decoding functions with
$P_{e}\rightarrow0$ as $n\rightarrow\infty$ \cite{Cover_book_2ndEdition}.
The capacity region of the GTRC is the convex closure of the set of
achievable rate pairs $\left(R_{1},R_{2}\right)$.

Note that the model depicted in Fig. \ref{fig:System-Model} is referred
to as the symmetric model if all source and relay nodes have the same
transmit powers, noise variances, i.e., $P=P_{R}$ and $N_{1}=N_{2}=N_{R}$.

\subsection{Lattice Definitions}

Here, we provide some necessary definitions on lattices and nested
lattice codes \cite{Erez_IT_04,Nazar_IT_11,conway_book}. 
\begin{defn}
(Lattice): An $n$-\textit{\textcolor{black}{dimensional lattice}}
$\Lambda$ is a set of points in Euclidean space $\mathbb{R}^{n}$
such that, if $\boldsymbol{x},\boldsymbol{y}\in\Lambda$, then $\boldsymbol{x}+\boldsymbol{y}\in\Lambda$,
and if $\boldsymbol{x}\in\Lambda$ , then $-\boldsymbol{x}\in\Lambda$.
A lattice $\Lambda$ can always be written in terms of a generator
matrix $\mathbf{G}\in\mathbb{Z}^{n\times n}$ as 
\[
\Lambda=\{\boldsymbol{x}=\boldsymbol{z}\mathbf{G}:\boldsymbol{z}\in\mathbb{Z}^{n}\},
\]
 where $\mathbb{Z}$ represents integers. 
\begin{defn}
(Quantizer): The \textit{\textcolor{black}{nearest neighbor quantizer}}
$\mathcal{Q}(.)$ associated with the lattice $\Lambda$ is 
\[
\mathcal{Q}_{\Lambda}(\boldsymbol{x})=\arg\underset{\boldsymbol{l}\in\Lambda}{\min}\left\Vert \boldsymbol{x}-\boldsymbol{l}\right\Vert .
\]

\begin{defn}
(Voronoi Region): The \textit{\textcolor{black}{fundamental Voronoi
region}} of a lattice $\Lambda$ is set of points in $\mathbb{R}^{n}$
closest to the zero codeword, i.e., 
\end{defn}
\end{defn}
\end{defn}
\[
\mathcal{V}_{0}(\Lambda)=\{\boldsymbol{x}\in\mathbb{\mathbb{R}}^{n}:\mathcal{Q}(\boldsymbol{x})=0\}.
\]

\begin{defn}
(Moments): $\sigma^{2}\left(\Lambda\right)$ which is called the second
moment of lattice $\Lambda$ is given by 
\begin{equation}
\sigma^{2}(\Lambda)=\frac{1}{n}\frac{\int_{\mathcal{V}(\Lambda)}\left\Vert \boldsymbol{x}\right\Vert ^{2}d\boldsymbol{x}}{\int_{\mathcal{V}(\Lambda)}d\boldsymbol{x}}.\label{eq:SM}
\end{equation}
 and the \textit{\textcolor{black}{normalized second moment}} of lattice
$\Lambda$ is 
\[
G(\Lambda)=\frac{\sigma^{2}(\Lambda)}{[\int_{\mathcal{V}(\Lambda)}d\boldsymbol{x}]^{\frac{2}{n}}}=\frac{\sigma^{2}(\Lambda)}{V^{\frac{2}{n}}},
\]
 where $V=\int_{\mathcal{V}(\Lambda)}d\boldsymbol{x}$ is the Voronoi
region volume, i.e., $V=\mbox{ Vol}(\mathcal{V})$. 
\begin{defn}
(Modulus): The \textit{\textcolor{black}{modulo-}}$\Lambda$ \textit{\textcolor{black}{operation}}
with respect to lattice $\Lambda$ is defined as 
\[
\boldsymbol{x}\mbox{ mod }\Lambda=\boldsymbol{x}-\mathcal{Q}(\boldsymbol{x}),
\]

\end{defn}
\end{defn}
that maps $\boldsymbol{x}$ into a point in the fundamental Voronoi
region. 
\begin{defn}
(Quantization Goodness or Rogers-good): A sequence of lattices $\Lambda^{(n)}\subseteq\mathbb{R}^{n}$
is good for mean-squared error (MSE) quantization if 
\[
\underset{n\rightarrow\infty}{\lim}G\left(\Lambda^{(n)}\right)=\frac{1}{2\pi e}.
\]

\end{defn}
The sequence is indexed by the lattice dimension $n$. The existence
of such lattices is shown in \cite{Zamir_IT_96,Erez_IT_05}. 
\begin{defn}
\label{(AWGN-channel-coding} (AWGN channel coding goodness or Poltyrev-good):
Let $\boldsymbol{Z}$ be a length-$i.i.d$ Gaussian vector, $\boldsymbol{Z}\thicksim\mathcal{N}\left(\boldsymbol{0},\sigma_{Z}^{2}\boldsymbol{I}_{n}\right)$.
The volume-to-noise ratio of a lattice is given by 
\[
\mu\left(\Lambda,\epsilon\right)=\frac{\left(\mbox{ Vol}(\mathcal{V})\right)^{2/n}}{\sigma_{Z}^{2}},
\]
 where $\sigma_{Z}^{2}$ is chosen such that $\mbox{ Pr}\left\{ \boldsymbol{Z}\notin\mathcal{V}\right\} =\epsilon$
and $\boldsymbol{I}_{n}$ is an $n\times n$ identity matrix. A sequence
of lattices is $\Lambda^{(n)}$ Poltyrev-good if 
\[
\underset{n\rightarrow\infty}{\lim}\mu\left(\Lambda^{(n)},\epsilon\right)=2\pi e,\,\,\,\,\,\forall\epsilon\in\left(0,1\right)
\]
 and, for fixed volume-to-noise ratio greater than $2\pi e$, $\mbox{ Pr}\left\{ \boldsymbol{Z}\notin\mathcal{V}^{n}\right\} $
decays exponentially in $n$ . 
\end{defn}
Poltyrev showed that sequences of such lattices exist \cite{Poltyrev_IT_94}.
The existence of a sequence of lattices $\Lambda^{(n)}$ which are
good in both senses (i.e., simultaneously are Poltyrev-good and Rogers-good)
has been shown in \cite{Erez_IT_05}. 
\begin{defn}
(Nested Lattices): A lattice $\Lambda^{(n)}$ is said to be nested
in lattice $\Lambda_{1}^{(n)}$ if $\Lambda^{(n)}\subseteq\Lambda_{1}^{(n)}$.
$\Lambda^{(n)}$ is referred to as the coarse lattice and $\Lambda_{1}^{(n)}$
as the fine lattice.

(Nested Lattice Codes): A nested lattice code is the set of all points
of a fine lattice $\Lambda_{1}^{(n)}$ that are within the fundamental
Voronoi region $\mathcal{V}$ of a coarse lattice $\Lambda^{(n)}$,
\[
\mathcal{C}=\left\{ \Lambda_{1}\cap\mathcal{V}\right\} .
\]
 The rate of a nested lattice code is 
\[
R=\frac{1}{n}\log\left|\mathcal{C}\right|=\frac{1}{n}\log\frac{\mbox{ Vol}\left(\mathcal{V}\right)}{\mbox{ Vol}\left(\mathcal{V}_{1}\right)}.
\]
 The existence of nested lattices where the coarse lattice as well
as the fine lattice are good in both senses has also been shown in
\cite{Nazar_IT_11,Krithivasan_arxiv}. An interesting property of
these codes is that any integer combinations of transmitted codewords
are themselves codewords. 
\end{defn}
In the following, we present a key property of dithered nested lattice
codes. 
\begin{lem}
\textbf{The Crypto Lemma \cite{Forney_Allerton_2003,Erez_IT_04}}
Let $\mathbf{V}$ be a random vector with an arbitrary distribution
over $\mathbb{R}^{n}$. If $\boldsymbol{D}$ is independent of $\mathbf{V}$
and uniformly distributed over $\mathcal{V}$, then $(\mathbf{V}+\boldsymbol{D})\textrm{ mod }\Lambda$
is also independent of $\mathbf{V}$ and uniformly distributed over
$\mathcal{V}$. 
\begin{IEEEproof}
See Lemma 2 in \textbf{\cite{Forney_Allerton_2003}.} 
\end{IEEEproof}
\end{lem}

\section{\label{sec:Lattice-based-achievable-rate} Nested Lattice Codes}

\subsection{Relay Strategies}

In this Section, we introduce two strategies for processing and transmitting
at the relay. In both schemes, we recover a linear combination of
messages instead of separate recovery of messages at the relay.

\subsubsection{Scheme 1: Compute-and-Forward}

The compute-and-forward strategy is proposed in \cite{Nazar_IT_11}.
In this scheme, the goal is recover an integer linear combination
of codewords, $\boldsymbol{V}_{1}$ and $\boldsymbol{V}_{2}$, i.e.,
we estimate 
\begin{equation}
\boldsymbol{V}_{1}+a\boldsymbol{V}_{2},\label{eq:linearCombMessages}
\end{equation}
 where $a\in\mathbb{Z}$. Since the transmitted sequences are from
lattice codes, it guarantee that any integer linear combination of
the codewords is a codeword. However, at the receiver, the received
signal which is a linear combination of the transmitted codewords
is no longer integer since the channel coefficients are real (or complex).
Also, the received signal is corrupted by noise. As a solution, Nazer
and Gastpar \cite{Nazar_IT_11} propose to scale the received signal
by a factor such that the obtained vector is made as close as possible
to an integer linear combination of the transmitted codewords.

To reach to this goal, for $n$ large enough, we assume that there
exist three lattices $\Lambda_{1}^{(n)}$, $\Lambda_{2}^{(n)}$ and
$\Lambda^{(n)}$ such that $\Lambda^{(n)}\subseteq\Lambda_{1}^{(n)}\subseteq\Lambda_{2}^{(n)}$.
$\Lambda_{1}^{(n)}$ and $\Lambda_{2}^{(n)}$ are both Poltyrev-good
and Rogers- good and $\Lambda^{(n)}$ is Rogers-good with the second
moment 
\[
\sigma^{2}\left(\Lambda\right)=P.
\]
 We denote the Voronoi regions of $\Lambda_{1}^{(n)}$, $\Lambda_{2}^{(n)}$
and $\Lambda^{(n)}$ with $\mathcal{V}_{1}$, $\mathcal{V}_{2}$ and
$\mathcal{V}$, respectively.

Encoding: We choose two codebooks $\mathcal{C}_{1}$ and $\mathcal{C}_{2}$,
such that 
\begin{eqnarray*}
\mathcal{C}_{1} & = & \left\{ \Lambda_{1}\cap\mathcal{V}\right\} ,\\
\mathcal{C}_{2} & = & \left\{ \Lambda_{2}\cap\mathcal{V}\right\} .
\end{eqnarray*}
 Now, for each input node $i$, the message set $\left\{ 1,...,2^{nR_{i}}\right\} $
is arbitrarily one-to-one mapped onto $\mathcal{C}_{i}$. We also
define two random dither vectors $\boldsymbol{D}_{i}\sim{\rm {Unif}\left(\mathcal{V}\right)}$
for $i=1,2$. Dither vectors are independent of each other and also
independent of the message of each node and the noise. Dither $\boldsymbol{D}_{i}$
is known to both the input nodes and the relay node. To transmit a
message, node $i$ chooses $\boldsymbol{V}_{i}\in\mathcal{C}_{i}$
associated with the message and sends 
\[
\boldsymbol{X}_{i}=\left[\boldsymbol{V}_{i}-\boldsymbol{D}_{i}\right]\mbox{ mod }\Lambda.
\]
 Note that by the crypto lemma, $\boldsymbol{X}_{i}$ is uniformly
distributed over $\mathcal{V}$ and independent of $\boldsymbol{V}_{i}$.
Thus, the average transmit power at node is equal to $\sigma^{2}\left(\Lambda\right)=P$,
so that the power constraint is satisfied.

Decoding: Upon receiving $\boldsymbol{Y}_{R}$, the relay node computes
\begin{eqnarray*}
\boldsymbol{Y}_{d_{R}} & = & \left[\alpha\boldsymbol{Y}_{R}+\boldsymbol{D}_{1}+a\boldsymbol{D}_{2}\right]\textrm{ mod }\Lambda,\\
 & = & \left[\alpha\boldsymbol{X}_{1}+\alpha\sqrt{g}\boldsymbol{X}_{2}+\alpha\boldsymbol{Z}_{R}+\boldsymbol{D}_{1}+a\boldsymbol{D}_{2}\right]\textrm{ mod }\Lambda,\\
 & \overset{\left(a\right)}{=} & \left[\boldsymbol{V}_{1}+a\boldsymbol{V}_{2}+\alpha\boldsymbol{X}_{1}-\left(\boldsymbol{V}_{1}-\boldsymbol{D}_{1}\right)+\alpha\sqrt{g}\boldsymbol{X}_{2}-a\left(\boldsymbol{V}_{2}-\boldsymbol{D}_{2}\right)+\alpha\boldsymbol{Z}_{R}\right]\textrm{ mod }\Lambda,\\
 & \overset{\left(b\right)}{=} & \left[\boldsymbol{V}_{1}+a\boldsymbol{V}_{2}+(\alpha-1)\boldsymbol{X}_{1}+(\alpha\sqrt{g}-a)\boldsymbol{X}_{2}+\alpha\boldsymbol{Z}_{R}\right]\textrm{ mod }\Lambda,\\
 & = & \left[\boldsymbol{T}+\boldsymbol{Z}_{{\rm eff}}\right]\textrm{ mod }\Lambda,
\end{eqnarray*}
 where (a) follows by adding and subtracting the term $\boldsymbol{V}_{1}+a\boldsymbol{V}_{2}$,
and (b) follows from the distributive law of modulo-$\Lambda$ operation
\cite{Zamir_09_ITA,Nazar_IT_11}, i.e., 
\[
\left[\left[\boldsymbol{X}\mbox{ mod }\Lambda\right]+\boldsymbol{Y}\right]\mbox{ mod }\Lambda=\left[\boldsymbol{X}+\boldsymbol{Y}\right]\mbox{ mod }\Lambda,
\]
 $\boldsymbol{X},\boldsymbol{Y}\in\mathbb{R}^{n}$. The effective
noise is given by 
\[
\boldsymbol{Z}_{{\rm eff}}=(\alpha-1)\boldsymbol{X}_{1}+(\alpha\sqrt{g}-a)\boldsymbol{X}_{2}+\alpha\boldsymbol{Z}_{R},
\]
 and 
\[
\boldsymbol{T}=\left[\boldsymbol{V}_{1}+a\boldsymbol{V}_{2}\right]\textrm{ mod }\Lambda.
\]
 Since $\boldsymbol{V}_{1},\boldsymbol{V}_{2},\boldsymbol{X}_{1},\boldsymbol{X}_{2}$
are independent, and also independent of $\boldsymbol{Z}_{R}$, and
using Crypto lemma, $\boldsymbol{Z}_{{\rm eff}}$ is independent of
$\boldsymbol{V}_{1}$ and $\boldsymbol{V}_{2}$ and thus independent
of $\boldsymbol{T}$. The relay aims to recover $\boldsymbol{T}$
from $\boldsymbol{Y}_{d_{R}}$ instead of recovering $\boldsymbol{V}_{1}$
and $\boldsymbol{V}_{2}$ individually. Due to the lattice chain,
i.e., $\Lambda^{(n)}\subseteq\Lambda_{1}^{(n)}\subseteq\Lambda_{2}^{(n)}$,
$\boldsymbol{T}$ is a point from $\Lambda_{2}^{n}$. To get an estimate
of $\boldsymbol{T}$, this vector is quantized onto $\Lambda_{2}^{n}$
modulo the lattice $\Lambda^{n}$: 
\begin{eqnarray*}
\hat{\boldsymbol{T}} & = & \mathcal{Q}_{\Lambda_{2}}\left(\boldsymbol{Y}_{d_{R}}\right)\mbox{ mod }\Lambda,\\
 & = & \mathcal{Q}_{\Lambda_{2}}\left(\boldsymbol{T}+\boldsymbol{Z}_{{\rm eff}}\right)\mbox{ mod }\Lambda,
\end{eqnarray*}
 where $\mathcal{Q}_{\Lambda_{2}}$ denotes the nearest neighbor lattice
quantizer associated with $\Lambda_{2}$. Thus, the decoding error
probability at the relay vanishes as $n\rightarrow\infty$ if 
\begin{equation}
\textrm{ Pr}\left(\boldsymbol{Z}_{{\rm eff}}\notin\mathcal{V}_{2}\right)\rightarrow0.\label{eq:Error Probibility}
\end{equation}
 By using Lemma 8 in \cite{Nazar_IT_11}, we know that the density
of $\boldsymbol{Z}_{{\rm eff}}$ can be upper bounded by the density
of $\boldsymbol{Z}_{{\rm eff}}^{*}$, which is an $i.i.d$ zero-mean
Gaussian vector whose variance approaches 
\[
N_{eq}=\left(\left(\alpha\sqrt{g}-a\right)^{2}+\left(\alpha-1\right)^{2}\right)P+\alpha^{2}N_{R},
\]
 as $n\rightarrow\infty$. From Definition \ref{(AWGN-channel-coding},
this means that $\textrm{ Pr}\left(\boldsymbol{Z}_{{\rm eff}}^{*}\notin\mathcal{V}_{2}\right)\rightarrow0$
so long as the volume-to-noise ratio satisfies 
\[
\mu\left(\Lambda_{2}\right)=\frac{\left(\mbox{ Vol}\left(\mathcal{V}_{2}\right)\right)^{2/n}}{N_{eq}}>2\pi e.
\]
 Therefore, for the volume of each Voronoi region, we have: 
\begin{equation}
\mbox{ Vol}\left(\mathcal{V}_{i}\right)>\left(2\pi eN_{eq}\right)^{n/2}\,\,\,\,\,\, i=1,2.\label{eq:Condition on the Volume of fine lattices}
\end{equation}
 For the volume of the fundamental Voronoi region of $\Lambda^{(n)}$,
we have: 
\begin{equation}
\mbox{ Vol}\left(\mathcal{V}\right)=\left(\frac{P}{G\left(\Lambda\right)}\right)^{n/2}.\label{eq:Condition on the Volume of Coarse  lattices}
\end{equation}
 Now, by using (\ref{eq:Condition on the Volume of fine lattices})
and (\ref{eq:Condition on the Volume of Coarse  lattices}) and definition
of the rate of a nested lattice code, we can achieve the following
rate for each node: 
\[
R_{i}<\frac{1}{2}\log\left(\frac{P}{G\left(\Lambda\right)2\pi eN_{eq}}\right),\,\,\,\,\,\, i=1,2.
\]
 Since $\Lambda$ is Rogers-good, $G\left(\Lambda\right)2\pi e\rightarrow1$
as $n\rightarrow\infty$. Thus, 
\begin{equation}
R_{i}<\frac{1}{2}\log\left(\frac{P}{\left(\left(\alpha\sqrt{g}-a\right)^{2}+\left(\alpha-1\right)^{2}\right)P+\alpha^{2}N_{R}}\right),\,\,\,\,\,\, i=1,2.\label{eq:Compute and Forward rate}
\end{equation}
 Now, we choose $\alpha$ as the minimum mean-square error (MMSE)
coefficient that minimizes the variance of the effective noise, $N_{eq}$.
Thus, we get 
\begin{equation}
\alpha_{_{{\rm MMSE}}}=\frac{\left(a\sqrt{g}+1\right)P}{\left(g+1\right)P+N_{R}}.\label{eq:Optimum alpha}
\end{equation}
 By inserting (\ref{eq:Compute and Forward rate}) in (\ref{eq:Optimum alpha}),
we can achieve any rate satisfying 
\begin{equation}
R_{i}<\frac{1}{2}\log\left(\frac{P}{\left(\left(\alpha_{_{{\rm MMSE}}}\sqrt{g}-a\right)^{2}+\left(\alpha_{_{{\rm MMSE}}}-1\right)^{2}\right)P+\alpha_{_{{\rm MMSE}}}^{2}N_{R}}\right).\label{eq:Scheme 1 rate}
\end{equation}
 i.e., it is possible to decode $\boldsymbol{T}$ within arbitrarily
low error probability, if the coding rates of the nested lattice codes
associated with the lattice partition $\Lambda^{(n)}\subseteq\Lambda_{1}^{(n)}$
and $\Lambda^{(n)}\subseteq\Lambda_{2}^{(n)}$ satisfy (\ref{eq:Scheme 1 rate}).
In this scheme, we must only obtain an integer linear combination
of messages. Since we want to obtain an estimate of $\boldsymbol{V}_{1}+a\boldsymbol{V}_{2}$,
$a$ should be an integer. On the other hand, we aim to obtain higher
achievable rates as much as we can. To reach this goal, we choose
$a$ as the closest integer to $\sqrt{g}$ i.e., $a=\left\lceil \sqrt{g}\right\rfloor $.

\subsubsection{Scheme 2: Our proposed scheme}

Let us first consider a theorem that is a key to our code construction. 
\begin{thm}
\label{thm:NAM-NestedLattices}\cite{Nam_IT_2011} For any $P_{1}\geq P_{2}\geq0$,
a sequence of $n$-dimensional lattice partition chains $\Lambda_{C}^{(n)}/\Lambda_{2}^{(n)}/\Lambda_{1}^{(n)}$,
i.e., $\Lambda_{1}^{(n)}\subseteq\Lambda_{2}^{(n)}\subseteq\Lambda_{C}^{(n)}$,
exists that satisfies the following properties:\end{thm}
\begin{itemize}
\item $\Lambda_{1}^{(n)}$ and $\Lambda_{2}^{(n)}$ are simultaneously Rogers-good
and Poltyrev- good while $\Lambda_{C}^{(n)}$ is Poltyrev-good. 
\item For any $\epsilon>0$ , $P_{i}-\epsilon\leq\sigma^{2}\left(\Lambda_{i}^{(n)}\right)\leq P_{i}$,
$i\in\{1,2\}$ for sufficiently large $n$. 
\item The coding rate of the nested lattice code associated with the lattice
partition $\Lambda_{C}^{(n)}/\Lambda_{1}^{(n)}$ is 
\[
R_{1}=\frac{1}{n}\log\left(|\mathcal{C}_{1}|\right)=\frac{1}{n}\log\left(\frac{\mbox{Vol}\left(\mathcal{V}_{1}\right)}{\mbox{Vol}\left(\mathcal{V}_{c}\right)}\right)=\gamma+o_{n}\left(1\right),
\]
 where $\mathcal{C}_{1}=\left\{ \Lambda_{C}^{(n)}\cap\mathcal{V}_{1}\right\} $
and $o_{n}\left(1\right)\rightarrow0$ as $n\rightarrow\infty$. The
coding rate of the nested lattice code associated with $\Lambda_{C}^{(n)}/\Lambda_{2}^{(n)}$
is given by 
\[
R_{2}=\frac{1}{n}\log\left(|\mathcal{C}_{2}|\right)=\frac{1}{n}\log\left(\frac{\mbox{ Vol}\left(\mathcal{V}_{2}\right)}{\mbox{ Vol}\left(\mathcal{V}_{c}\right)}\right)=R_{1}+\frac{1}{2}\log\left(\frac{P_{2}}{P_{1}}\right),
\]
 where $\mathcal{C}_{2}=\left\{ \Lambda_{C}^{(n)}\cap\mathcal{V}_{2}\right\} $. \end{itemize}
\begin{IEEEproof}
The proof of theorem is given in \cite{Nam_IT_2011}. 
\end{IEEEproof}
In the following, by applying a lattice-based coding scheme, we obtain
achievable rate region at the relay. Suppose that there exist three
lattices $\Lambda_{1}^{(n)}$, $\Lambda_{2}^{(n)}$ and $\Lambda_{3}^{(n)}=\sqrt{g}\Lambda_{2}^{(n)}$,
which are Rogers-good (i.e.,$\underset{n\rightarrow\infty}{\lim}G\left(\Lambda_{i}^{(n)}\right)=\frac{1}{2\pi e},\,\textrm{for\,\,\,}i=1,2,3\,)$,
and Poltyrev-good with the following second moments 
\[
\sigma^{2}\left(\Lambda_{1}\right)=P,\,\,\,\textrm{and}\,\,\sigma^{2}\left(\Lambda_{3}\right)=gP,
\]
 and a lattice $\Lambda_{C}^{(n)}$ which is Poltyrev-good with 
\begin{eqnarray*}
\Lambda_{1}^{(n)} & \subseteq & \Lambda_{C}^{(n)},\\
\Lambda_{3}^{(n)} & \subseteq & \Lambda_{C}^{(n)}.
\end{eqnarray*}
 Encoding: To transmit both messages, we construct the following codebooks:
\begin{eqnarray*}
\mathcal{C}_{1} & = & \left\{ \Lambda_{C}\cap\mathcal{V}_{1}\right\} ,\\
\mathcal{C}_{2} & = & \left\{ \frac{\Lambda_{C}}{\sqrt{g}}\cap\mathcal{V}_{2}\right\} .
\end{eqnarray*}
 Then node $i$ chooses $\boldsymbol{V}_{i}\in\mathcal{C}_{i}$ associated
with the message $W_{i}$ and sends 
\[
\boldsymbol{X}_{i}=\left[\boldsymbol{V}_{i}-\boldsymbol{D}_{i}\right]\textrm{ mod }\Lambda_{i},
\]
 where \textcolor{black}{$\boldsymbol{D}_{1}$} and \textcolor{black}{$\boldsymbol{D}_{2}$}
are t\textcolor{black}{wo independent dithers} that are uniformly
distributed over Voronoi regions $\mathcal{V}_{1}$ and $\mathcal{V}_{2}$,
respectively. Dithers are known at the source nodes and the relay.
Due to the crypto-lemma, $\boldsymbol{X}_{i}$ is uniformly distributed
over $\mathcal{V}_{i}$ and independent of $\boldsymbol{V}_{i}$.
Thus, the average transmit power of node $i$ is equal to $P$, and
the power constraint is met.

Decoding: At the relay node, based on the channel output that is given
by 
\begin{equation}
\boldsymbol{Y}_{R}=\boldsymbol{X}_{1}+\sqrt{g}\boldsymbol{X}_{2}+\boldsymbol{Z}_{R},\label{eq:input decoder}
\end{equation}
 we estimate $\boldsymbol{V}_{1}+\sqrt{g}\boldsymbol{V}_{2}$. Depend
on the value of $g$, we consider two cases:

\subsubsection*{Case (I): $g\geq1$}

Based on Theorem \ref{thm:NAM-NestedLattices}, we can find two lattices,
$\Lambda_{1}^{(n)}$ and $\Lambda_{3}^{(n)}$, such that $\Lambda_{3}^{(n)}\subseteq\Lambda_{1}^{(n)}$
. With this selection of lattices, the relay node performs the following
operation:

\begin{eqnarray}
\boldsymbol{Y}_{d_{R}} & = & \left[\alpha\boldsymbol{Y}_{R}+\boldsymbol{D}_{1}+\sqrt{g}\boldsymbol{D}_{2}\right]\textrm{ mod }\Lambda_{1}\nonumber \\
 & = & \left[\alpha\boldsymbol{X}_{1}+\alpha\sqrt{g}\boldsymbol{X}_{2}+\alpha\boldsymbol{Z}_{R}+\boldsymbol{D}_{1}+\sqrt{g}\boldsymbol{D}_{2}\right]\textrm{ mod }\Lambda_{1}\nonumber \\
 & = & \left[\boldsymbol{V}_{1}+\sqrt{g}\boldsymbol{V}_{2}+\alpha\boldsymbol{X}_{1}-\left(\boldsymbol{V}_{1}-\boldsymbol{D}_{1}\right)+\alpha\sqrt{g}\boldsymbol{X}_{2}-\sqrt{g}\left(\boldsymbol{V}_{2}-\boldsymbol{D}_{2}\right)+\alpha\boldsymbol{Z}_{R}\right]\textrm{ mod }\Lambda_{1}\nonumber \\
 & \overset{\left(\text{c}\right)}{=} & \left[\left[\boldsymbol{V}_{1}+\sqrt{g}\boldsymbol{V}_{2}\right]\textrm{ mod }\Lambda_{3}+(\alpha-1)\boldsymbol{X}_{1}+\sqrt{g}(\alpha-1)\boldsymbol{X}_{2}+\alpha\boldsymbol{Z}_{R}\right]\textrm{ mod }\Lambda_{1}\label{eq:Effective Noise g>1}\\
 & = & \left[\boldsymbol{T}+\boldsymbol{Z}_{{\rm eff}}\right]\textrm{ mod }\Lambda_{1},\nonumber 
\end{eqnarray}
 where (c) follows from $\Lambda_{3}^{(n)}\subseteq\Lambda_{1}^{(n)}$
and $\Lambda_{3}^{(n)}=\sqrt{g}\Lambda_{2}^{(n)}$ and the distributive
law of modulo-$\Lambda$ operation. The effective noise is given by
\[
\boldsymbol{Z}_{{\rm eff}}=\left[(\alpha-1)\boldsymbol{X}_{1}+\sqrt{g}(\alpha-1)\boldsymbol{X}_{2}+\alpha\boldsymbol{Z}_{R}\right]\textrm{ mod }\Lambda_{1},
\]
 and 
\[
\boldsymbol{T}=\left[\boldsymbol{V}_{1}+\sqrt{g}\boldsymbol{V}_{2}\right]\textrm{ mod }\Lambda_{3}.
\]
 Due to the dithers, the vectors $\boldsymbol{V}_{1},\boldsymbol{V}_{2},\boldsymbol{X}_{1},\boldsymbol{X}_{2}$
are independent, and also independent of $\boldsymbol{Z}_{R}$. Therefore,
$\boldsymbol{Z}_{{\rm eff}}$ is independent of $\boldsymbol{V}_{1}$
and $\boldsymbol{V}_{2}$. From the crypto-lemma, it follows that
$\boldsymbol{T}$ is uniformly distributed over $\sqrt{g}\mathcal{C}_{2}$
and independent of $\boldsymbol{Z}_{{\rm eff}}$ \cite{Nam_IT_2011}.

The problem of finding the optimum value for $\alpha$ when the lattice
dimension goes to infinity, reduces to obtain the value of $\alpha$
that minimizes the effective noise variance. Hence, by minimizing
variance of $\boldsymbol{Z}_{{\rm eff}}$, we obtain 
\begin{equation}
\alpha_{_{{\rm MMSE}}}=\frac{\left(g+1\right)P}{\left(g+1\right)P+N_{R}}.\label{eq:OptimalAlpha}
\end{equation}
 The relay attempts to recover $\boldsymbol{T}$ from $\boldsymbol{Y}_{d_{R}}$
instead of recovering $\boldsymbol{V}_{1}$ and $\boldsymbol{V}_{2}$
individually. The method of decoding is minimum Euclidean distance
lattice decoding \cite{Erez_IT_04,ElGamal_IT_04,Poltyrev_IT_94},
which finds the closest point to $\boldsymbol{Y}_{d_{R}}$ in $\Lambda_{C}^{(n)}$.
Thus, the estimate of $\boldsymbol{T}$ is given by, 
\[
\hat{\boldsymbol{T}}=\mathcal{Q}_{\Lambda_{C}}\left(\boldsymbol{Y}_{d_{R}}\right).
\]
 Then, from the type of decoding, the probability of decoding error
is given by 
\begin{eqnarray*}
P_{e} & = & \textrm{ Pr}\left\{ \hat{\boldsymbol{T}}\neq\boldsymbol{T}\right\} \\
 & = & \textrm{ Pr}\left\{ \boldsymbol{Z}_{{\rm eff}}\notin\mathcal{V}_{C}\right\} .
\end{eqnarray*}
 Now, we have the following theorem which bounds the error probability. 
\begin{thm}
\label{thm:For-lattice-partition}For the described lattice partition
chain and any rate $R_{1}$ satisfying 
\[
R_{1}<R_{1}^{*}=\left[\frac{1}{2}\log\left(\frac{1}{g+1}+\frac{P}{N_{R}}\right)\right]^{+},
\]
 the error probability under minimum Euclidean distance lattice decoding
is bounded by 
\[
P_{e}=e^{-n\left(E_{P}\left(2^{2\left(R_{1}^{*}-R_{1}\right)}\right)-o_{n}\left(1\right)\right)},
\]
 where $E_{p}\left(.\right)$ is the Poltyrev exponent, which is given
by \cite{Poltyrev_IT_94} 
\begin{eqnarray}
E_{P}\left(x\right) & = & \begin{cases}
\frac{x}{2}-\frac{1}{2}\left(x-1-\ln x\right), & 1\leq x\leq2\\
\frac{1}{2}\left(1+\ln\frac{x}{4}\right), & 2\leq x\leq4\\
\frac{x}{8}, & x\geq4
\end{cases}\label{eq:PoltyrevExponent}
\end{eqnarray}
 and $\left[x\right]^{+}=\max\left(0,x\right)$.\end{thm}
\begin{IEEEproof}
The proof of theorem is similar to the proof of theorem 3 in \cite{Nam_IT_2011}
and removed here. 
\end{IEEEproof}
Since $E_{P}\left(x\right)>0$ for $x>1$, the error probability vanishes
as $n\rightarrow\infty$ if $R_{1}<R_{1}^{*}$. Thus, by Theorem \ref{thm:NAM-NestedLattices}
and Theorem \ref{thm:For-lattice-partition}, the error probability
at the relay node vanishes if 
\begin{eqnarray}
R_{1} & \leq & \left[\frac{1}{2}\log\left(\frac{1}{g+1}+\frac{P}{N_{R}}\right)\right]^{+},\label{eq:achievable rate by scheme 2 R_1}\\
R_{2} & \leq & \left[\frac{1}{2}\log\left(\frac{g}{g+1}+\frac{gP}{N_{R}}\right)\right]^{+}.\label{eq:achievable rate by scheme 2 R_2}
\end{eqnarray}
 Clearly, using a time sharing argument the following rates can be
achieved: 
\begin{eqnarray*}
R_{1} & \leq & u.c.e\left\{ \left[\frac{1}{2}\log\left(\frac{1}{g+1}+\frac{P}{N_{R}}\right)\right]^{+}\right\} ,\\
R_{2} & \leq & u.c.e\left\{ \left[\frac{1}{2}\log\left(\frac{g}{g+1}+\frac{gP}{N_{R}}\right)\right]^{+}\right\} ,
\end{eqnarray*}
 where u.c.e is the upper convex envelope with respect to $\frac{P}{N_{R}}$
.

\begin{figure}
\begin{centering}
\includegraphics[width=12cm]{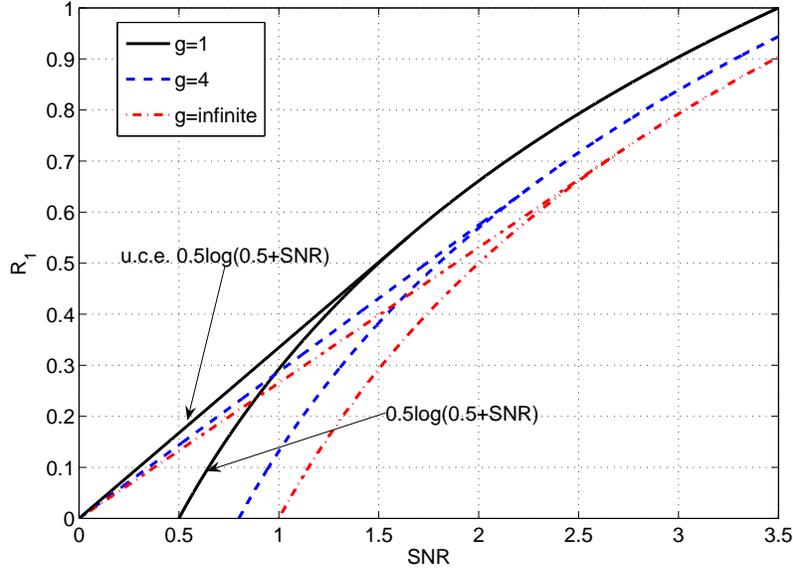} 
\par\end{centering}

\caption{\label{fig:Comoarsion-of-achievable}Comparison of the achievable
rate $R_{1}$ for different values of $g$.}
\end{figure}

At low SNR, i.e., ${\rm {SNR}}\leq\frac{g}{g+1}$, pure (infinite
dimensional) lattice-strategies cannot achieve any positive rates
for $R_{1}$ as shown in Fig. \ref{fig:Comoarsion-of-achievable}.
Hence, time sharing is required between the point ${\rm {SNR=0}}$
and ${\rm {SNR^{*}}}$, which is a solution of the following equation:
\[
f({SNR})=\frac{df({SNR})}{d{SNR}}{SNR},
\]
 where $f(x)=\frac{1}{2}\log\left(\frac{1}{g+1}+x\right)$. We also
evaluate numerically the achievable rates for $R_{1}$ with lattice
strategies for different values of $g$. As we observe, with increasing
$g$, the achievable rate with lattice scheme decreases. As it is
shown in Fig. \ref{fig:Comoarsion-of-achievable}, the maximum difference
between two extreme cases ($g=1$ and $g=\infty$) is 0.1218 bit.

\subsubsection*{Case (II): $g<1$}

By using Theorem \ref{thm:NAM-NestedLattices}, we can choose two
lattices $\Lambda_{1}^{(n)}$ and $\Lambda_{3}^{(n)}$ such that $\Lambda_{1}^{(n)}\subseteq\Lambda_{3}^{(n)}$.
The relay calculates $\boldsymbol{Y}_{d_{R}}=\left[\alpha\boldsymbol{Y}_{R}+\boldsymbol{D}_{1}+\sqrt{g}\boldsymbol{D}_{2}\right]\textrm{ mod }\Lambda_{3}$.
The equivalent channel is given by

\begin{eqnarray}
\boldsymbol{Y}_{d_{R}} & = & \left[\alpha\boldsymbol{X}_{1}+\alpha\sqrt{g}\boldsymbol{X}_{2}+\alpha\boldsymbol{Z}_{R}+\boldsymbol{D}_{1}+\sqrt{g}\boldsymbol{D}_{2}\right]\textrm{ mod }\Lambda_{3}\nonumber \\
 & = & \left[\boldsymbol{V}_{1}+\sqrt{g}\boldsymbol{V}_{2}+\alpha\boldsymbol{X}_{1}-\left(\boldsymbol{V}_{1}-\boldsymbol{D}_{1}\right)+\alpha\sqrt{g}\boldsymbol{X}_{2}-\sqrt{g}\left(\boldsymbol{V}_{2}-\boldsymbol{D}_{2}\right)+\alpha\boldsymbol{Z}_{R}\right]\textrm{ mod }\Lambda_{3}\nonumber \\
 & \overset{\left(d\right)}{=} & \left[\boldsymbol{V}_{1}+\sqrt{g}\boldsymbol{V}_{2}+(\alpha-1)\boldsymbol{X}_{1}+\sqrt{g}(\alpha-1)\boldsymbol{X}_{2}+\alpha\boldsymbol{Z}_{R}\right]\textrm{ mod }\Lambda_{3}\label{eq:Effective Noise g<1}\\
 & = & \left[\boldsymbol{T}+\boldsymbol{Z}_{{\rm eff}}\right]\textrm{ mod }\Lambda_{3},\nonumber 
\end{eqnarray}
 where (d) follows from $\Lambda_{1}^{(n)}\subseteq\Lambda_{3}^{(n)}$
and $\Lambda_{3}^{(n)}=\sqrt{g}\Lambda_{2}^{(n)}$ and the distributive
law of modulo-$\Lambda$ operation. The effective noise is given by
\[
\boldsymbol{Z}_{{\rm eff}}=\left[(\alpha-1)\boldsymbol{X}_{1}+\sqrt{g}(\alpha-1)\boldsymbol{X}_{2}+\alpha\boldsymbol{Z}_{R}\right]\textrm{ mod }\Lambda_{3},
\]
 and 
\[
\boldsymbol{T}=\left[\boldsymbol{V}_{1}+\sqrt{g}\boldsymbol{V}_{2}\right]\textrm{ mod }\Lambda_{1}.
\]
 Due to the dithers, the vectors $\boldsymbol{V}_{1},\boldsymbol{V}_{2},\boldsymbol{X}_{1},\boldsymbol{X}_{2}$
are independent, and also independent of $\boldsymbol{Z}_{R}$. Therefore,
$\boldsymbol{Z}_{{\rm eff}}$ is independent of $\boldsymbol{V}_{1}$
and $\boldsymbol{V}_{2}$. From the crypto-lemma, it follows that
$\boldsymbol{T}$ is uniformly distributed over $\mathcal{C}_{1}$
and independent of $\boldsymbol{Z}_{{\rm eff}}$ \cite{Nam_IT_2011}.
In order to achieve the maximal rate, the optimal MMSE factor is used,
i.e., 
\begin{equation}
\alpha=\alpha_{_{{\rm MMSE}}}=\frac{\left(g+1\right)P}{\left(g+1\right)P+N_{R}}.\label{eq:OptimalAlpha-1}
\end{equation}
 Similar to case $g\geq1$, instead of recovering $\boldsymbol{V}_{1}$
and $\boldsymbol{V}_{2}$ separately, the relay recovers $\boldsymbol{T}$.
Again, the decoding method is minimum Euclidean distance lattice decoding,
which finds the closest point to $\boldsymbol{Y}_{d_{R}}$ in $\Lambda_{C}^{(n)}$.
Thus, the estimate of $\boldsymbol{T}$ is given by, 
\[
\hat{\boldsymbol{T}}=\mathcal{Q}_{\Lambda_{C}}\left(\boldsymbol{Y}_{d_{R}}\right).
\]

\begin{thm}
\label{thm:For-lattice-partition-1}For the described lattice partition
chain, if 
\[
R_{2}<R_{2}^{*}=\left[\frac{1}{2}\log\left(\frac{g}{g+1}+\frac{gP}{N_{R}}\right)\right]^{+},
\]
 the error probability under minimum Euclidean distance lattice decoding
is bounded by 
\[
P_{e}=e^{-n\left(E_{P}\left(2^{2\left(R_{2}^{*}-R_{2}\right)}\right)-o_{n}\left(1\right)\right)},
\]
 where $E_{p}\left(.\right)$ is the Poltyrev exponent given in (\ref{eq:PoltyrevExponent}). \end{thm}
\begin{IEEEproof}
The proof of theorem is similar to the proof of theorem 3 in \cite{Nam_IT_2011}
and removed here. 
\end{IEEEproof}
Here also the error probability vanishes as $n\rightarrow\infty$
if $R_{2}<R_{2}^{*}$ since $E_{P}\left(x\right)>0$ for $x>1$. Thus,
by Theorem \ref{thm:NAM-NestedLattices} and Theorem \ref{thm:For-lattice-partition-1},
the error probability at the relay vanishes if 
\begin{eqnarray*}
R_{1} & \leq & \left[\frac{1}{2}\log\left(\frac{1}{g+1}+\frac{P}{N_{R}}\right)\right]^{+},\\
R_{2} & \leq & \left[\frac{1}{2}\log\left(\frac{g}{g+1}+\frac{gP}{N_{R}}\right)\right]^{+},
\end{eqnarray*}
 Clearly, using a time sharing argument the following rates can be
achieved: 
\begin{eqnarray}
R_{1} & \leq & u.c.e\left\{ \left[\frac{1}{2}\log\left(\frac{1}{g+1}+\frac{P}{N_{R}}\right)\right]^{+}\right\} ,\label{eq:achievable Symmetric rate by lattice and time sharing}\\
R_{2} & \leq & u.c.e\left\{ \left[\frac{1}{2}\log\left(\frac{g}{g+1}+\frac{gP}{N_{R}}\right)\right]^{+}\right\} ,\label{eq:achievable rate R_2 by scheme 2}
\end{eqnarray}
 where u.c.e is the upper convex envelope with respect to $\frac{P}{N}$.

\begin{figure}
\begin{centering}
\includegraphics[width=12cm]{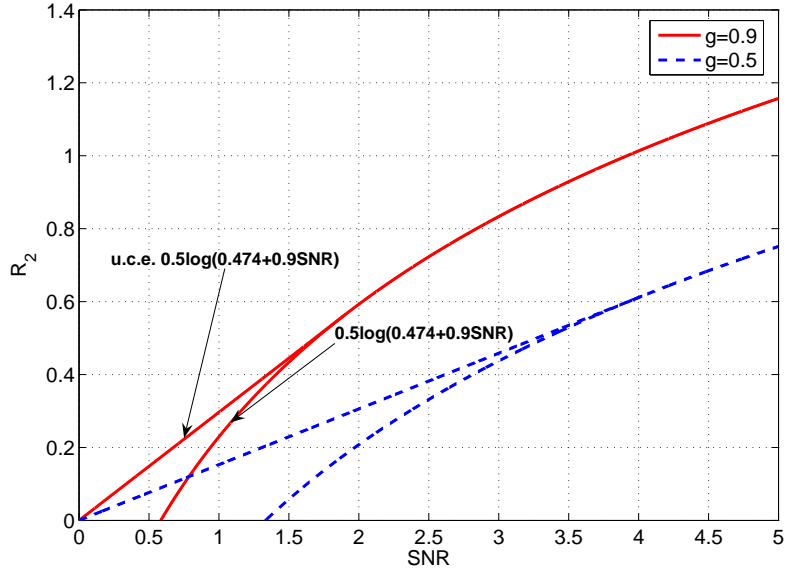} 
\par\end{centering}

\caption{\label{fig:Comoarsion-of-achievable-1}Comparison of the achievable
symmetric rate for different values of $g$}
\end{figure}

For ${\rm {SNR}\leq\frac{1}{g\left(g+1\right)}}$, pure (infinite
dimensional) lattice-strategies cannot achieve any positive rate for
$R_{2}$ as shown in Fig. \ref{fig:Comoarsion-of-achievable-1}. Hence,
time sharing is required between the point ${\rm {SNR}=0}$ and ${\rm {SNR^{*}}}$,
which is a solution of the following: 
\[
f({SNR})=\frac{df({SNR})}{d{SNR}}{SNR},
\]
 where $f(x)=\frac{1}{2}\log\left(\frac{g}{g+1}+gx\right)$. We also
evaluate numerically the achievable rate $R_{2}$ of lattice strategy
for different values of $g$. As we see with decreasing $g$, the
achievable rate with lattice scheme is decreased.

\subsection{Broadcast Phase}

We assume that the relay can recover the linear combination of both
messages correctly, i.e., \textcolor{black}{there is no error in the
MAC phase, $\hat{\boldsymbol{T}}=\boldsymbol{T}$. The relay attempts
to broadcast a message such that each node can recover the other node's
message based on both the received signal from the relay node and
the available side information at each node, i.e., its own message.
For the decoding at node 1 and node 2, we can use jointly typical
decoding or lattice based scheme. Here, we apply jointly typical decoding.
We consider scheme 2; decoding for scheme 1 is similar to scheme 2.
For scheme 2, we also assume that $g\geq1$. Under this assumption,
we have $R_{2}\geq R_{1}$. Now, we generate $2^{nR_{2}}$-sequences
with each element $i.i.d$. according to $\mathcal{N}\sim\left(0,P_{R}\right)$.
These sequences form a codebook $\mathcal{C}_{R}$. We assume that
there is a one-to-one correspondence between $\mathcal{C}_{2}$ and
$\mathcal{C}_{R}$. }

\textcolor{black}{Let us denote the relay codeword by $\boldsymbol{X}_{R}\left(\boldsymbol{T^{'}}\right)$.
Based on $\boldsymbol{Y}_{2}=\boldsymbol{X}_{R}+\boldsymbol{Z}_{2}$,
node 2 estimates the relay message $\boldsymbol{T^{'}}$ as $\hat{\boldsymbol{T}_{2}^{'}}$
if a unique codeword $\boldsymbol{X}_{R}\left(\hat{\boldsymbol{T}_{2}^{'}}\right)\in\mathcal{C}_{R,2}$
exists such that $\left(\boldsymbol{X}_{R}\left(\hat{\boldsymbol{T}_{2}^{'}}\right),\boldsymbol{Y}_{2}\right)$
are jointly typical, where 
\[
\mathcal{C}_{R,2}=\left\{ \boldsymbol{X}_{R}\left(\boldsymbol{T^{'}}\right):\boldsymbol{T^{'}}=\left[\boldsymbol{V}_{1}+\sqrt{g}\boldsymbol{v}_{2}\right]\textrm{ mod }\Lambda_{3},\boldsymbol{v}_{2}\in\mathcal{C}_{2}\right\} .
\]
 Note that $|\mathcal{C}_{R,2}|=2^{nR_{1}}$. Now, by using the knowledge
of $\boldsymbol{V}_{2}$ and $\hat{\boldsymbol{T}_{2}^{'}}$, node
2 estimates the message of node 1 as: 
\[
\hat{\boldsymbol{V}_{1}}=\left[\hat{\boldsymbol{T}}_{2}-\sqrt{g}\boldsymbol{v}_{2}\right]\textrm{ mod }\Lambda_{3}.
\]
 From the argument of random coding and jointly typical decoding \cite{Cover_book_2ndEdition},
we get 
\begin{equation}
R_{1}\leq\frac{1}{2}\log\left(1+\frac{P_{R}}{N_{2}}\right).\label{eq:Rate at the node 2}
\end{equation}
 Similarly, at node 1, we get } 
\begin{equation}
R_{2}\leq\frac{1}{2}\log\left(1+\frac{P_{R}}{N_{1}}\right).\label{eq:Rate at the node 1}
\end{equation}
 Now, we summarize our results for both schemes in the following two
theorem: 
\begin{thm}
\label{thm:For-Gaussian-two-way}For the Gaussian two-way relay channel,
the following rate region is achievable: 
\begin{eqnarray}
R_{1} & \leq & \min\left(\frac{1}{2}\log\left(\frac{P}{\left(\left(\alpha_{_{{\rm MMSE}}}\sqrt{g}-a\right)^{2}+\left(\alpha_{_{{\rm MMSE}}}-1\right)^{2}\right)P+\alpha_{_{{\rm MMSE}}}^{2}N_{R}}\right),\frac{1}{2}\log\left(1+\frac{P_{R}}{N_{2}}\right)\right),\label{eq:Rate region for Scheme 1}\\
R_{2} & \leq & \min\left(\frac{1}{2}\log\left(\frac{P}{\left(\left(\alpha_{_{{\rm MMSE}}}\sqrt{g}-a\right)^{2}+\left(\alpha_{_{{\rm MMSE}}}-1\right)^{2}\right)P+\alpha_{_{{\rm MMSE}}}^{2}N_{R}}\right),\frac{1}{2}\log\left(1+\frac{P_{R}}{N_{1}}\right)\right),\label{eq:Rate region for scheme 1 R_2}
\end{eqnarray}
 where $\alpha_{_{{\rm MMSE}}}=\frac{\left(a\sqrt{g}+1\right)P}{\left(g+1\right)P+N_{R}}$
and $a$ is the closest integer to $\sqrt{g}$ i.e., $a=\left\lceil \sqrt{g}\right\rfloor $.\end{thm}
\begin{IEEEproof}
The proof of theorem follows from the achievable rate-region of scheme
1 at the relay (\ref{eq:Scheme 1 rate}) and achievable rates at nodes
1 and 2 (\ref{eq:Rate at the node 2}), (\ref{eq:Rate at the node 1}). \end{IEEEproof}
\begin{thm}
\label{thm:For-Gaussian-two-way-1}For Gaussian two-way relay channel,
the following rate region is achievable: 
\begin{eqnarray}
R_{1} & \leq & \min\left(u.c.e\left\{ \left[\frac{1}{2}\log\left(\frac{1}{g+1}+\frac{P}{N_{R}}\right)\right]^{+}\right\} ,\frac{1}{2}\log\left(1+\frac{P_{R}}{N_{2}}\right)\right),\label{eq:Rate region for Scheme 2}\\
R_{2} & \leq & \min\left(u.c.e\left\{ \left[\frac{1}{2}\log\left(\frac{g}{g+1}+\frac{gP}{N_{R}}\right)\right]^{+}\right\} ,\frac{1}{2}\log\left(1+\frac{P_{R}}{N_{1}}\right)\right).\label{eq:Rate region for scheme 2 R_2}
\end{eqnarray}
 \end{thm}
\begin{IEEEproof}
The proof of theorem follows from the achievable rate-region of scheme
2 at the relay (\ref{eq:achievable Symmetric rate by lattice and time sharing}),(\ref{eq:achievable rate R_2 by scheme 2})
and achievable rates at nodes 1 and 2 (\ref{eq:Rate at the node 2}),
(\ref{eq:Rate at the node 1}). 
\end{IEEEproof}

\section{\label{sec:Outer-Bound}Outer Bound }

By using the cut-set bound, an outer bound for a TRC can be derived.
If a rate pair $\left(R_{1},R_{2}\right)$ is achievable for a general
TRC, then 
\begin{eqnarray}
R_{1} & \leq & \min\left\{ I\left(\boldsymbol{X}_{1};\boldsymbol{Y}_{R},\boldsymbol{Y}_{2}|\boldsymbol{X}_{R},\boldsymbol{X}_{2}\right),I\left(\boldsymbol{X}_{1},\boldsymbol{X}_{R};\boldsymbol{Y}_{2}|,\boldsymbol{X}_{2}\right)\right\} ,\label{eq:Cutset bound for R_1}\\
R_{2} & \leq & \min\left\{ I\left(\boldsymbol{X}_{2};\boldsymbol{Y}_{R},\boldsymbol{Y}_{1}|\boldsymbol{X}_{R},\boldsymbol{X}_{1}\right),I\left(\boldsymbol{X}_{2},\boldsymbol{X}_{R};\boldsymbol{Y}_{1}|,\boldsymbol{X}_{1}\right)\right\} ,\label{eq:Cutset bound for R_2}
\end{eqnarray}
 where the minimization is over a joint probability $p\left(x_{1},x_{2},x_{R}\right)$.
If we evaluate the outer bound for the GTRC, all the terms will be
minimized by the product distribution $p\left(x_{1},x_{2},x_{R}\right)=p\left(x_{1}\right)p\left(x_{2}\right)p\left(x_{R}\right)$,
where $p\left(x_{1}\right)$, $p\left(x_{2}\right)$ and $p\left(x_{R}\right)$
are Gaussian pdfs with zero means and variances of $P_{1}$, $P_{2}$
and $P_{R}$, respectively. Therefore, from (\ref{eq:Cutset bound for R_1})
and (\ref{eq:Cutset bound for R_2}), one can get \cite{Knopp_GDR_2007}
\begin{eqnarray}
R_{1} & \leq & \min\left(\frac{1}{2}\log\left(1+\frac{P}{N_{R}}\right),\frac{1}{2}\log\left(1+\frac{P_{R}}{N_{2}}\right)\right),\label{eq:Cutset bound for Gaussian TRC}\\
R_{2} & \leq & \min\left(\frac{1}{2}\log\left(1+\frac{gP}{N_{R}}\right),\frac{1}{2}\log\left(1+\frac{P_{R}}{N_{1}}\right)\right).\label{eq:Cutset bound for Gaussian TRC-1}
\end{eqnarray}
 Although there is a gap between the outer region and the achievable
rate-region, we show that the gap vanishes at high SNR and hence the
capacity region is completely determined in this limit of high SNR. 
\begin{cor}
\label{cor:At-high-SNR,}At high SNR, the capacity region of the GTRC,
is given by 
\begin{eqnarray}
R_{1} & \leq & \min\left(\frac{1}{2}\log\left(1+\frac{P}{N_{R}}\right),\frac{1}{2}\log\left(1+\frac{P_{R}}{N_{2}}\right)\right)-o(1),\label{eq:Capacity at high SNR}\\
R_{2} & \leq & \min\left(\frac{1}{2}\log\left(1+\frac{gP}{N_{R}}\right),\frac{1}{2}\log\left(1+\frac{P_{R}}{N_{1}}\right)\right)-o(1),\label{eq:Capacity at high SNR R_2}
\end{eqnarray}
 where $o(1)\rightarrow0$ as $SNR\rightarrow\infty$.\end{cor}
\begin{IEEEproof}
Using (\ref{eq:Effective Noise g>1}) and (\ref{eq:Effective Noise g<1})
with $\alpha=1$ and taking $\Lambda_{1}$ and $\Lambda_{3}$ to be
lattices that are Poltyrev-good and Rogers-good with the second moment
equals to $P$ and $gP$, respectively, we get the following achievable
rate region at the relay: 
\begin{eqnarray*}
R_{1} & \leq & \frac{1}{2}\log\left(\frac{P}{N}\right),\\
R_{2} & \leq & \frac{1}{2}\log\left(\frac{gP}{N}\right).
\end{eqnarray*}
 Now, by using (\ref{eq:Rate at the node 2}) and (\ref{eq:Rate at the node 1}),
we can achieve the following rate region for GTRC: 
\begin{eqnarray}
R_{1} & \leq & \min\left(\frac{1}{2}\log\left(\frac{P}{N}\right),\frac{1}{2}\log\left(1+\frac{P_{R}}{N_{2}}\right)\right),\label{eq:Rate region at the high SNR(R_1)}\\
R_{2} & \leq & \min\left(\frac{1}{2}\log\left(\frac{gP}{N}\right),\frac{1}{2}\log\left(1+\frac{P_{R}}{N_{1}}\right)\right).\label{eq:Rate region at high SNR(R_2)}
\end{eqnarray}
 By evaluating the outer bound in (\ref{eq:Cutset bound for Gaussian TRC}),
and (\ref{eq:Cutset bound for Gaussian TRC-1}) for ${\rm {SNR}}\rightarrow\infty$,
we observe that we can approach the outer bound at high SNR and the
proof is complete. 
\end{IEEEproof}

\subsection{Calculating the gap}

In the following, we bound the gap between the outer region, given
in (\ref{eq:Cutset bound for Gaussian TRC}), (\ref{eq:Cutset bound for Gaussian TRC-1})
and the achievable region in (\ref{eq:Rate region for Scheme 2}),
(\ref{eq:Rate region for scheme 1 R_2}). \textcolor{black}{It is
easy to show that the gap for the symmetric model, i.e., $P_{R}=P$
and $N_{1}=N_{2}=N_{R}\overset{\triangle}{=}N$ is always greater
than the gap in other cases.} Thus, we focus on the symmetric model.
For the symmetric model, the outer bound on the capacity region is
given by: 
\begin{eqnarray}
R_{1} & \leq & \frac{1}{2}\log\left(1+\frac{P}{N}\right),\label{eq:Outer bound for symmetric TRC}\\
R_{2} & \leq & \min\left(\frac{1}{2}\log\left(1+\frac{gP}{N}\right),\frac{1}{2}\log\left(1+\frac{P}{N}\right)\right),\label{eq:Outer bound for symmetric TRC R_2}
\end{eqnarray}
 and the achievable rate region is simplified as: 
\begin{eqnarray}
R_{1} & \leq & u.c.e\left\{ \left[\frac{1}{2}\log\left(\frac{1}{g+1}+\frac{P}{N}\right)\right]^{+}\right\} ,\label{eq: achiebale rate region for symmetric model  with scheme 1}\\
R_{2} & \leq & \min\left(u.c.e\left\{ \left[\frac{1}{2}\log\left(\frac{g}{g+1}+\frac{gP}{N}\right)\right]^{+}\right\} ,\frac{1}{2}\log\left(1+\frac{P}{N}\right)\right).\label{eq:achievable rate region for symmetric model with scheme 1 R_2}
\end{eqnarray}
 To obtain the gap between the outer bound and the achievable region,
first we bound the gap for $R_{1}$.

\subsubsection{Gap for $R_{1}$}

Let us define this gap, denoted by $\xi\left(P,g,N\right)$, as the
following: 
\[
\xi\left(P,g,N\right)\overset{\triangle}{=}\frac{1}{2}\log\left(1+\frac{P}{N}\right)-u.c.e\left\{ \left[\frac{1}{2}\log\left(\frac{1}{g+1}+\frac{P}{N}\right)\right]^{+}\right\} .
\]
 In the following theorem, we provide an upper bound on $\xi\left(P,g,N\right)$. 
\begin{thm}
\label{thm:GAP-FOR-R1}Let $x^{*}$be the solution of the equation
$\frac{x}{x+\frac{1}{g+1}}=\log_{e}\left(x+\frac{1}{g+1}\right)$.
For any $P,g,N$ , the gap is bounded by $\xi\left(P,g,N\right)$
\begin{equation}
\xi\left(P,g,N\right)\leq\max\left\{ \frac{1}{2}\log\left(x^{*}+\frac{1}{g+1}\right)-\frac{x^{*}-\frac{g}{g+1}}{2\ln2\left(x^{*}+\frac{1}{g+1}\right)},\frac{1}{2}\log\left(1+\frac{\frac{g}{g+1}}{x^{*}+\frac{1}{g+1}}\right)\right\} .\label{eq:Gap for R_1}
\end{equation}
 \end{thm}
\begin{IEEEproof}
Let us define $x\overset{\triangle}{=}\frac{P}{N}$. 
\[
\xi\left(P,g,N\right)=\frac{1}{2}\log\left(1+x\right)-u.c.e\left\{ \left[\frac{1}{2}\log\left(x+\frac{1}{g+1}\right)\right]^{+}\right\} \overset{\Delta}{=}\tilde{\xi}\left(x\right).
\]
 Now, we calculate the upper convex envelope with respect to $x$.
First, we compute the line through the origin which is tangent to
$\frac{1}{2}\log\left(x+\frac{1}{g+1}\right)$: 
\[
y=\frac{x}{2\ln2\left(x^{*}+\frac{1}{g+1}\right)},
\]
 where $x^{*}$is a solution of the following equation 
\[
\frac{x}{x+\frac{1}{g+1}}=\log_{e}\left(x+\frac{1}{g+1}\right).
\]
 Therefore, we get 
\[
u.c.e\left\{ \left[\frac{1}{2}\log\left(\frac{1}{g+1}+\frac{P}{N}\right)\right]^{+}\right\} =\begin{cases}
\frac{1}{2}\log\left(x+\frac{1}{g+1}\right) & x\geq x^{*}\\
\frac{x}{2\ln2\left(x^{*}+\frac{1}{g+1}\right)} & 0\leq x\leq x^{*}
\end{cases},
\]

a) For $x\geq x^{*}$: $\tilde{\xi}\left(x\right)$ is given by 
\begin{equation}
\tilde{\xi}\left(x\right)=\frac{1}{2}\log\left(\frac{x+1}{x+\frac{1}{g+1}}\right)=\frac{1}{2}\log\left(1+\frac{\frac{g}{g+1}}{x+\frac{1}{g+1}}\right).\label{eq:up1}
\end{equation}
 Since $\tilde{\xi}\left(x\right)$ is decreasing with respect to
$x$, hence $\tilde{\xi}\left(x\right)$ is maximized for $x=x^{*}$.

b) For $0\leq x\leq x^{*}$: $\tilde{\xi}\left(x\right)$ is given
by 
\[
\tilde{\xi}\left(x\right)=\frac{1}{2}\log\left(1+x\right)-\frac{x}{2\ln2\left(x^{*}+\frac{1}{g+1}\right)}.
\]
 The maximum of $\tilde{\xi}\left(x\right)$ occurs at $x_{m}=x^{*}-\frac{g}{g+1}$,
hence we get 
\begin{equation}
\tilde{\xi}\left(x\right)\leq\tilde{\xi}\left(x_{m}\right)=\frac{1}{2}\log\left(x^{*}+\frac{1}{g+1}\right)-\frac{x^{*}-\frac{g}{g+1}}{2\ln2\left(x^{*}+\frac{1}{g+1}\right)}.\label{eq:up2}
\end{equation}
 Combining (\ref{eq:up1}) and (\ref{eq:up2}) completes the proof. 
\end{IEEEproof}

\subsubsection{Gap for $R_{2}$}

To obtain the gap for $R_{2}$, we consider two cases:

\subsubsection*{Case 1: $g<1$}

In this case, the gap between the outer bound (\ref{eq:Outer bound for symmetric TRC R_2})
and the achievable rate in (\ref{eq:achievable rate region for symmetric model with scheme 1 R_2})
is defined as the following: 
\[
\eta_{1}\left(P,g,N\right)=\frac{1}{2}\log\left(1+\frac{gP}{N}\right)-u.c.e\left\{ \left[\frac{1}{2}\log\left(\frac{g}{g+1}+\frac{gP}{N}\right)\right]^{+}\right\} .
\]

\begin{thm}
\label{thm:GAP-FOR-R2}Let $x^{*}$ be the solution of the equation
$\frac{x}{x+\frac{1}{g+1}}=\log_{e}\left(gx+\frac{g}{g+1}\right)$.
For any $P,g,N$, the gap for $R_{2}$ is bounded by 
\begin{equation}
\eta_{1}\left(P,g,N\right)\leq\max\left\{ \frac{1}{2}\log\left(gx^{*}+\frac{g}{g+1}\right)-\frac{gx^{*}-\frac{1}{\left(g+1\right)}}{2\ln2\left(gx^{*}+\frac{g}{g+1}\right)},\frac{1}{2}\log\left(\frac{gx^{*}+1}{gx^{*}+\frac{g}{g+1}}\right)\right\} .\label{eq:Gap for g<1 R_2}
\end{equation}
 \end{thm}
\begin{IEEEproof}
Let $x\overset{\triangle}{=}\frac{P}{N}$. Then the gap is 
\[
\eta_{1}\left(P,g,N\right)=\frac{1}{2}\log\left(gx+1\right)-u.c.e\left\{ \left[\frac{1}{2}\log\left(gx+\frac{g}{g+1}\right)\right]^{+}\right\} \overset{\Delta}{=}\tilde{\eta}_{1}\left(x\right).
\]
 Here, again we calculate the upper convex envelope. In a similar
way, we obtain

\[
u.c.e\left\{ \left[\frac{1}{2}\log\left(gx+\frac{g}{g+1}\right)\right]^{+}\right\} =\begin{cases}
\frac{1}{2}\log\left(gx+\frac{g}{g+1}\right) & x\geq x^{*}\\
\frac{x}{2\ln2\left(x^{*}+\frac{1}{g+1}\right)} & 0\leq x\leq x^{*}
\end{cases},
\]
 where $x^{*}$is the solution of the following equation 
\[
\frac{x}{x+\frac{1}{g+1}}=\log_{e}\left(gx+\frac{g}{g+1}\right).
\]
 Now, we calculate $\tilde{\eta}_{1}\left(x\right)$ as follows:

a) For $x\geq x^{*}$: $\tilde{\eta}_{1}\left(x\right)$ is given
by 
\[
\tilde{\eta}_{1}\left(x\right)=\frac{1}{2}\log\left(\frac{gx+1}{gx+\frac{g}{g+1}}\right)=\frac{1}{2}\log\left(1+\frac{\frac{1}{g+1}}{gx+\frac{g}{g+1}}\right).
\]
 Since $\tilde{\eta}_{1}\left(x\right)$ is decreasing with $x$,
hence $\tilde{\eta}_{1}\left(x\right)$ is maximized at $x=x^{*}$.

b) For $0\leq x\leq x^{*}$: $\tilde{\eta}_{1}\left(x\right)$ is
given by 
\[
\tilde{\eta}_{1}\left(x\right)=\frac{1}{2}\log\left(gx+1\right)-\frac{x}{2\ln2\left(x^{*}+\frac{1}{g+1}\right)}.
\]
 The maximum of $\tilde{\eta}_{1}\left(x\right)$ occurs at $x^{*}-\frac{1}{g\left(g+1\right)}$,
hence, we get 
\[
\tilde{\eta}_{1}\left(x\right)\leq\tilde{\eta}_{1}\left(x^{*}-\frac{1}{g\left(g+1\right)}\right)=\frac{1}{2}\log\left(gx^{*}+\frac{g}{g+1}\right)-\frac{gx^{*}-\frac{1}{\left(g+1\right)}}{2\ln2\left(gx^{*}+\frac{g}{g+1}\right)}.
\]

\end{IEEEproof}

\subsubsection*{Case II: $g>1$}

Let $x^{*}$ be the solution of the equation $\frac{x}{x+\frac{1}{g+1}}=\log_{e}\left(gx+\frac{g}{g+1}\right)$.
Then, if there is no intersection between the straight line $y=\frac{x}{2\ln2\left(x^{*}+\frac{1}{g+1}\right)}$
and the curve\textcolor{red}{{} }$\frac{1}{2}\log\left(1+\frac{P}{N}\right)$,
the outer bound in (\ref{eq:Outer bound for symmetric TRC}) meet
the achievable rate $R_{2}$ in (\ref{eq: achiebale rate region for symmetric model  with scheme 1}).
Thus, the gap is zero. In other wise, we evaluate the gap as follows:
\begin{equation}
\eta_{2}\left(P,g,N\right)\overset{\triangle}{=}\frac{1}{2}\log\left(1+\frac{P}{N}\right)-u.c.e\left\{ \left[\frac{1}{2}\log\left(\frac{g}{g+1}+\frac{gP}{N}\right)\right]^{+}\right\} .\label{eq:GAPGHIGH}
\end{equation}
 If SNR is fixed, then $\eta_{2}\left(P,g,N\right)$ is a decreasing
function of $g$; thus $\eta_{2}\left(P,g,N\right)$ is maximized
at $g=1$. In the following theorem, we bound $\eta_{2}\left(P,g,N\right)$
for $g=1$. 
\begin{thm}
\label{thm:Let--be-2}Let $x^{*}$ be the solution of the equation
$\frac{x}{x+\frac{1}{2}}=\log_{e}\left(x+\frac{1}{2}\right)$. For
any $P$ and $N$, the gap for $R_{2}$ is bounded by 
\[
\eta_{2}\left(P,1,N\right)\leq0.167
\]
 \end{thm}
\begin{IEEEproof}
Let $x\overset{\triangle}{=}\frac{P}{N}$. We can rewrite (\ref{eq:GAPGHIGH})
as 
\[
\eta_{2}\left(P,g,N\right)=\frac{1}{2}\log\left(1+x\right)-u.c.e\left\{ \left[\frac{1}{2}\log\left(\frac{1}{2}+x\right)\right]^{+}\right\} \overset{\Delta}{=}\tilde{\eta}_{2}\left(x\right).
\]

a) For $x\geq x^{*}$: $\tilde{\eta}_{2}\left(x\right)$ is given
by 
\[
\tilde{\eta}_{2}\left(x\right)=\frac{1}{2}\log\left(\frac{1+x}{\frac{1}{2}+x}\right).
\]
 Since $\tilde{\eta}_{2}\left(x\right)$ is decreasing with respect
to $x$, $\tilde{\eta}_{2}\left(x\right)$ is maximized at $x=x^{*}\cong1.655$.

b) For $0\leq x\leq x^{*}$: $\tilde{\eta}_{2}\left(x\right)$ is
given by 
\[
\tilde{\eta}_{2}\left(x\right)=\frac{1}{2}\log\left(1+x\right)-\frac{x}{2\ln2\left(x^{*}+\frac{1}{2}\right)}.
\]
 The maximum of $\tilde{\eta}_{2}\left(x\right)$ occurs at $x^{*}-\frac{1}{2}$,
hence, we get 
\[
\tilde{\eta}_{2}\left(x\right)\leq\tilde{\eta}_{2}\left(x^{*}-\frac{1}{2}\right).
\]
 The Theorem holds since $\tilde{\eta}_{2}\left(x^{*}\right)\leq\tilde{\eta}_{2}\left(x^{*}-\frac{1}{2}\right)$. 
\end{IEEEproof}

\subsubsection{Maximum gap for $R_{1}$, $R_{2}$ and $R_{1}+R_{2}$}
\begin{thm}
\label{thm:MaxGapGLow} For $g<1$, the outer bound in (\ref{eq:Outer bound for symmetric TRC})
is within 0.2654 bit from the achievable rate in (\ref{eq: achiebale rate region for symmetric model  with scheme 1})
for user 1. In addition, the outer bound of (\ref{eq:Outer bound for symmetric TRC R_2})
is within 0.2658 bit from the achievable rate in (\ref{eq:achievable rate region for symmetric model with scheme 1 R_2})
for user 2 .\end{thm}
\begin{IEEEproof}
The gap $\xi\left(P,g,N\right)$ in (\ref{eq:Gap for R_1}) is an
increasing function of $g$, hence this gap is maximized for $g=\infty$.
Thus, the maximum of gap for $R_{1}$ is 0.2654 bit. For $g<1$, the
gap in (\ref{eq:Gap for g<1 R_2}) is decreasing with respect to $g$.
Therefore, this gap is maximized as $g\rightarrow0$. By evaluating
(\ref{eq:Gap for g<1 R_2}) for $g\rightarrow0$, we observe that
the maximum gap is 0.2658 bit. \end{IEEEproof}
\begin{thm}
\label{thm:MaxGapGHigh}For $g>1$, the outer bound in (\ref{eq:Outer bound for symmetric TRC})
is within 0.2654 bit from the achievable rate, given (\ref{eq: achiebale rate region for symmetric model  with scheme 1})
for user 1. For user 2, the outer bound in (\ref{eq:Outer bound for symmetric TRC R_2})
is within 0.167 bit from the achievable rate in (\ref{eq:achievable rate region for symmetric model with scheme 1 R_2}).\end{thm}
\begin{IEEEproof}
The gap in (\ref{eq:Gap for R_1}) is an increasing function with
respect to $g$; hence this gap is maximized for $g=\infty$. Thus,
the maximum of the gap for $R_{1}$ is 0.2654 bit. Now, by using Theorem
\ref{thm:Let--be-2}, we observe that the maximum of the gap for $R_{2}$
is 0.167 bit.\end{IEEEproof}
\begin{thm}
\label{thm:GapSumRateGLow}For $g<1$, the sum-rate gap of our proposed
scheme (i.e., scheme 2) is within 0.334 of the outer bound. \end{thm}
\begin{IEEEproof}
We have

\[
\zeta(P,g,N)=(R_{1}+R_{2})_{O}-(R_{1}+R_{2})_{I}=(R_{1O}-R_{1I})+(R_{2O}-R_{2I}),
\]
 By using (\ref{eq:Gap for R_1}) and (\ref{eq:Gap for g<1 R_2}),
we get

\begin{eqnarray}
\zeta_{1}(P,g,N) & \leq & \left\{ \max\left\{ \frac{1}{2}\log\left(x^{*}+\frac{1}{g+1}\right)-\frac{x^{*}-\frac{g}{g+1}}{2\ln2\left(x^{*}+\frac{1}{g+1}\right)},\frac{1}{2}\log\left(1+\frac{\frac{g}{g+1}}{x^{*}+\frac{1}{g+1}}\right)\right\} +\right.\nonumber \\
 &  & \left.\max\left\{ \frac{1}{2}\log\left(gy^{*}+\frac{g}{g+1}\right)-\frac{gy^{*}-\frac{1}{\left(g+1\right)}}{2\ln2\left(gy^{*}+\frac{g}{g+1}\right)},\frac{1}{2}\log\left(\frac{gy^{*}+1}{gy^{*}+\frac{g}{g+1}}\right)\right\} \right\} ,\label{eq:Gap for sum rate}
\end{eqnarray}
 where $x^{*}$is the solution of the equation $\frac{x}{x+\frac{1}{g+1}}=\log_{e}\left(x+\frac{1}{g+1}\right)$
and $y^{*}$ is the solution of the equation $\frac{y}{y+\frac{1}{g+1}}=\log_{e}\left(gy+\frac{g}{g+1}\right)$.
Now, by evaluating the gap (\ref{eq:Gap for sum rate}), which is
an increasing function with respect to $g$, we observe that the maximum
gap is 0.334 bit (For more detail, see Section \ref{sec:Numerical-Result}).\end{IEEEproof}
\begin{thm}
\label{thm:thm:GapSumRateGHigh}For $g>1$, the sum-rate gap of our
proposed scheme (i.e., scheme 2) is within 0.334 of the outer bound. \end{thm}
\begin{IEEEproof}
For the sum-rate gap, we have

\[
\zeta_{2}(P,g,N)=(R_{1}+R_{2})_{O}-(R_{1}+R_{2})_{I}=(R_{1O}-R_{1I})+(R_{2O}-R_{2I}).
\]
 The first term, using (\ref{eq:Gap for R_1}), ca be calculated as

\begin{equation}
R_{1O}-R_{1I}\leq\max\left\{ \frac{1}{2}\log\left(x^{*}+\frac{1}{g+1}\right)-\frac{x^{*}-\frac{g}{g+1}}{2\ln2\left(x^{*}+\frac{1}{g+1}\right)},\frac{1}{2}\log\left(1+\frac{\frac{g}{g+1}}{x^{*}+\frac{1}{g+1}}\right)\right\} ,\label{eq:Gap for fist term}
\end{equation}
 where $x^{*}$ is a solution of the equation $\frac{x}{x+\frac{1}{g+1}}=\log_{e}\left(x+\frac{1}{g+1}\right)$.
For evaluating the second term, we consider two cases:

1) There is no intersection between the straight line $y=\frac{x}{2\ln2\left(x^{*}+\frac{1}{g+1}\right)}$
and the curve\textcolor{red}{{} }$\frac{1}{2}\log\left(1+\frac{P}{N}\right)$,
where $x^{*}$ is the solution of the equation $\frac{x}{x+\frac{1}{g+1}}=\log_{e}\left(gx+\frac{g}{g+1}\right)$.
For this case, the outer bound (\ref{eq:Outer bound for symmetric TRC})
meet the achievable rate $R_{2}$ in (\ref{eq: achiebale rate region for symmetric model  with scheme 1}).
Thus, the gap for the second term is zero and it is sufficient to
calculate the gap for the first term. The gap in (\ref{eq:Gap for fist term})
is increasing with $g$ and is maximized for $g=\infty$. With evaluating
(\ref{eq:Gap for fist term}) for $g=\infty$, we observe that the
maximum gap is $0.2654$ bit.

2) If there is an intersection between the straight line $y=\frac{x}{2\ln2\left(x^{*}+\frac{1}{g+1}\right)}$
and the curve\textcolor{red}{{} }$\frac{1}{2}\log\left(1+\frac{P}{N}\right)$,
where $x^{*}$ is the solution of the equation $\frac{x}{x+\frac{1}{g+1}}=\log_{e}\left(gx+\frac{g}{g+1}\right)$,
we evaluate the following gap:

\[
\xi(P,g,N)\leq\max\left\{ \frac{1}{2}\log\left(x^{*}+\frac{1}{g+1}\right)-\frac{x^{*}-\frac{g}{g+1}}{2\ln2\left(x^{*}+\frac{1}{g+1}\right)},\frac{1}{2}\log\left(1+\frac{\frac{g}{g+1}}{x^{*}+\frac{1}{g+1}}\right)\right\} +\xi^{'}(P,g,N),
\]
 where 
\[
\xi^{'}\left(P,g,N\right)=\frac{1}{2}\log\left(1+\frac{P}{N}\right)-u.c.e\left\{ \left[\frac{1}{2}\log\left(\frac{g}{g+1}+\frac{gP}{N}\right)\right]^{+}\right\} .
\]
 The gap is decreasing with $g$. Thus, it is maximized for $g=1$.
By evaluating the gap for $g=1$, we get

\[
\xi(P,g,N)\leq0.334.
\]

\end{IEEEproof}

\section{\label{sec:Numerical-Result}Discussion and Numerical Result}

\begin{figure}
\begin{centering}
\includegraphics[width=12cm]{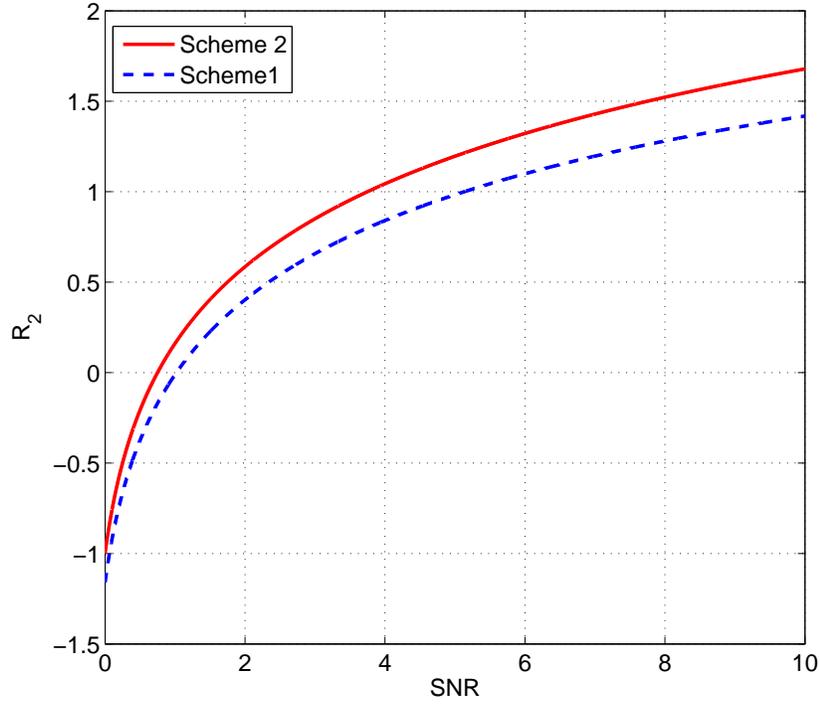} 
\par\end{centering}

\caption{\label{fig:Comparison-of-achievablefig3}Comparison of the achievable
rates of scheme 1 and scheme 2 for user 1, $R_{2}$, in the symmetric
GTRC.\textcolor{black}{{} The channel gain is $g=3$.}}
\end{figure}

\begin{figure}
\begin{centering}
\includegraphics[width=12cm]{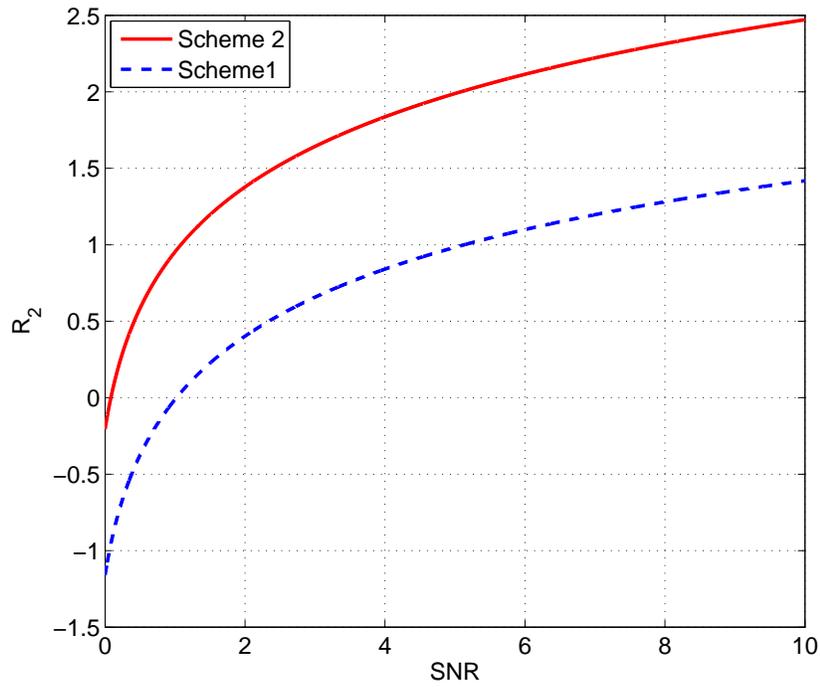} 
\par\end{centering}

\caption{\label{fig:Comparison-of-achievablefig4}Comparison of the achievable
rates of scheme 1 and scheme 2 for user 2, $R_{2}$, in the symmetric
GTRC. \textcolor{black}{The channel gain is $g=10$.}}
\end{figure}

First, we assume that\textcolor{black}{{} $\sqrt{g}$ }is an integer
number. Then, the achievable rate region of scheme 1 can be calculated
as: 
\begin{eqnarray*}
R_{1} & < & \left[\frac{1}{2}\log\left(\frac{1}{g+1}+\frac{P}{N_{R}}\right)\right]^{+}\\
R_{2} & < & \left[\frac{1}{2}\log\left(\frac{1}{g+1}+\frac{P}{N_{R}}\right)\right]^{+},
\end{eqnarray*}
 and the achievable rate region of scheme 2 is given by:

\begin{eqnarray*}
R_{1} & \leq & \left[\frac{1}{2}\log\left(\frac{1}{g+1}+\frac{P}{N}\right)\right]^{+},\\
R_{2} & \leq & \left[\frac{1}{2}\log\left(\frac{g}{g+1}+\frac{gP}{N}\right)\right]^{+}.
\end{eqnarray*}
 As we see, if $g$ is an integer number and larger than one, the
achievable rate region of scheme 2 is larger than that of scheme 2.
But for $g=1$, both schemes have the same performance as the proposed
scheme in \cite{Wilson_IT_2010}. For other \textcolor{black}{integer}
values of $g$ i.e., $g=2,3,4,...$, scheme 2 have better performance
than scheme 1. This comparison is depicted in Fig.\ref{fig:Comparison-of-achievablefig3}
for\textcolor{black}{{} $g=3$ }and in Fig.\ref{fig:Comparison-of-achievablefig4}
for $g=10.$ 

\begin{figure}
\begin{centering}
\includegraphics[width=12cm]{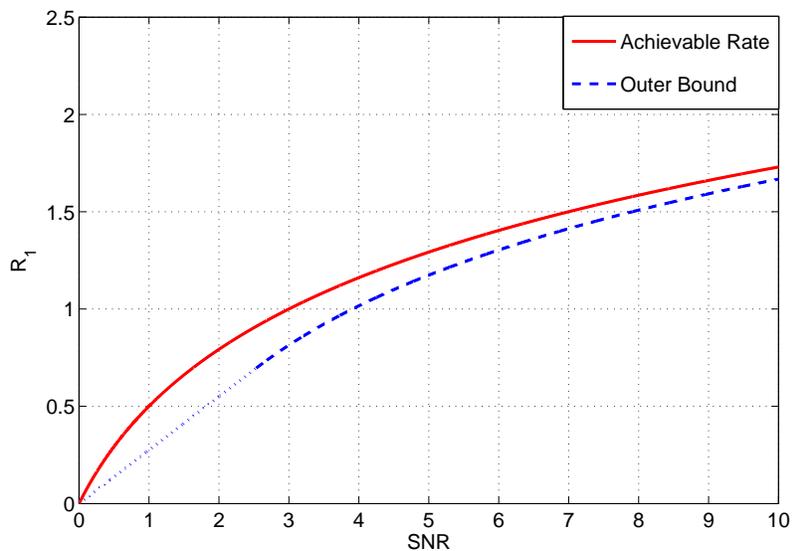} 
\par\end{centering}

\caption{\label{fig:Comparison-of-achievable-1}Comparison of the achievable
rate of scheme 2 for $R_{1}$ and the outer bound in the symmetric
GTRC. The channel gain is $g=10$.}
\end{figure}

In Table \ref{tab:GapUser1GHigh} and Table \ref{tab:GapUser1GLow},
we provide the gap between the achievable rate of user 1 $(R_{1})$
and the outer bound, given in (\ref{eq:Gap for R_1}). We observe
that when channel gain increases, the gap increases and the maximum
gap is 0.2654 bit. For $g=10$, we compare the achievable rate $(R_{1})$
of scheme 2 and the outer bound in Fig.\ref{fig:Comparison-of-achievable-1}.
As SNR increases, the gap decreases.

In Table \ref{tab:GapUser2} and for $g<1$, we compare the gap between
the achievable rate for user 2 $(R_{2})$ and the outer bound, given
in (\ref{eq:Gap for g<1 R_2}). As we observe, when the channel gain
decreases, the gap increases and the maximum gap is 0.2658 bit.

\begin{table}[H]
\caption{{\small \label{tab:GapUser1GHigh}The gap between the outer bound
and the achievable rate for user 1, $R_{1}$} for $g>1$}

\centering{}%
\begin{tabular}{|c|c|c|c|c|c|c|c|c|}
\hline 
Channel Gain $\left(g\right)$  & 1  & 4  & 9  & 16  & 25  & 64  & 100  & inf\tabularnewline
\hline 
Gap (bits)  & 0.167  & 0.2361  & 0.2497  & 0.2547  & 0.2573  & 0.2621  & 0.2637  & 0.2654\tabularnewline
\hline 
\end{tabular}
\end{table}

\begin{table}[H]
\caption{{\small \label{tab:GapUser1GLow}The gap between the outer bound and
the achievable rate for user 1, $R_{1}$} for $g<1$}

\centering{}%
\begin{tabular}{|c|c|c|c|c|c|c|c|c|}
\hline 
Channel Gain $\left(g\right)$  & 1  & 0.7  & 0.6  & 0.5  & 0.3  & 0.1  & 0.001  & 0.0001\tabularnewline
\hline 
Gap (bits)  & 0.167  & 0.146  & 0.136  & 0.1252  & 0.09497  & 0.045  & 0.000686  & 0.0001611\tabularnewline
\hline 
\end{tabular}
\end{table}

\begin{table}[H]
\caption{{\small \label{tab:GapUser2}The gap between the outer bound and the
achievable rate for user 2, $R_{2}$}}

\centering{}%
\begin{tabular}{|c|c|c|c|c|c|c|c|c|}
\hline 
Channel Gain $\left(g\right)$  & 1  & 0.7  & 0.6  & 0.5  & 0.3  & 0.1  & 0.001  & 0.0001\tabularnewline
\hline 
Gap (bits)  & 0.167  & 0.1872  & 0.195  & 0.2038  & 0.224  & 0.2498  & 0.2651  & 0.2658\tabularnewline
\hline 
\end{tabular}
\end{table}

\section{\label{sec:Conclusion}Conclusion}

In this paper, a Gaussian two way relay channel (GTRC) is considered.
By using nested lattice-based coding scheme, we obtain two achievable
rate regions for this channel and characterize the theoretical gap
between the achievable rate region and the outer bound. We show that
the gap is less than 0.2658 bit for each user, which is best gap-to-capacity
result to date. We also show that, at high SNR, the lattice based
coding scheme can achieve the capacity region.

 \bibliographystyle{IEEEtran}
\bibliography{IEEEabrv,ReferencesGhasemiSep2012}

\end{document}